\documentclass[a4paper,twocolumn,11pt,accepted=2026-04-10]{quantumarticle}
\pdfoutput=1
\usepackage[utf8]{inputenc}
\usepackage[english]{babel}
\usepackage[T1]{fontenc}
\usepackage{amsmath}
\usepackage{hyperref}

\usepackage{xcolor}
\usepackage{amssymb}
\usepackage{amsmath}
\usepackage{graphicx}
\usepackage{float}
\usepackage{bbm}
\usepackage{color}
\usepackage{comment}
\usepackage{gensymb}
\usepackage{lipsum}
\usepackage{todonotes}
\usepackage{siunitx}
\definecolor{blue}{rgb}{0,0,1}
\definecolor{grey}{rgb}{0.6,0.6,0.6}

\usepackage{tikz}
\usepackage{lipsum}

\begin{document}

\title{Non-Markovian thermal reservoirs for autonomous entanglement distribution}

\author{Joan Agustí}
\affiliation{Technical University of Munich, TUM School of Natural Sciences, Physics Department, 85748 Garching, Germany}
\affiliation{Walther-Meißner-Institut, Bayerische Akademie der Wissenschaften, 85748 Garching, Germany}
\affiliation{Munich Center for Quantum Science and Technology (MCQST), 80799 Munich, Germany}
\affiliation{Institute of Fundamental Physics IFF-CSIC, Calle Serrano 113b, 28006 Madrid, Spain} 
\orcid{0000-0002-9883-1958}
\email{joan.agusti@iff.csic.es}
\author{Christian M. F. Schneider}
\affiliation{Technical University of Munich, TUM School of Natural Sciences, Physics Department, 85748 Garching, Germany}
\affiliation{Walther-Meißner-Institut, Bayerische Akademie der Wissenschaften, 85748 Garching, Germany}
\affiliation{Munich Center for Quantum Science and Technology (MCQST), 80799 Munich, Germany}
\orcid{0000-0002-5766-7979}
\author{Kirill G. Fedorov}
\affiliation{Technical University of Munich, TUM School of Natural Sciences, Physics Department, 85748 Garching, Germany}
\affiliation{Walther-Meißner-Institut, Bayerische Akademie der Wissenschaften, 85748 Garching, Germany}
\affiliation{Munich Center for Quantum Science and Technology (MCQST), 80799 Munich, Germany}
\orcid{0000-0002-3243-4343}
\author{Stefan Filipp}
\affiliation{Technical University of Munich, TUM School of Natural Sciences, Physics Department, 85748 Garching, Germany}
\affiliation{Walther-Meißner-Institut, Bayerische Akademie der Wissenschaften, 85748 Garching, Germany}
\affiliation{Munich Center for Quantum Science and Technology (MCQST), 80799 Munich, Germany}
\orcid{0000-0002-1976-1817}
\author{Peter Rabl}
\affiliation{Technical University of Munich, TUM School of Natural Sciences, Physics Department, 85748 Garching, Germany}
\affiliation{Walther-Meißner-Institut, Bayerische Akademie der Wissenschaften, 85748 Garching, Germany}
\affiliation{Munich Center for Quantum Science and Technology (MCQST), 80799 Munich, Germany}
\orcid{0000-0002-2560-8835}
\maketitle

\begin{abstract}
We describe a novel scheme for the generation of stationary entanglement between two separated qubits that are driven by a purely thermal photon source. While in this scenario the qubits remain in a separable state at all times when the source is broadband, i.e. Markovian, the qubits relax into an entangled steady state once the bandwidth of the thermal source is sufficiently reduced. We explain this phenomenon by the appearance of a quasiadiabatic dark state and identify the most relevant nonadiabatic corrections that eventually lead to a breakdown of the entangled state, once the temperature is too high. This effect demonstrates how the non-Markovianity of an otherwise incoherent reservoir can be harnessed for quantum communication applications in optical, microwave, and phononic networks. As two specific examples, we discuss the use of filtered room-temperature noise as a passive resource for entangling distant superconducting qubits in a cryogenic quantum link or solid-state spin qubits in a phononic quantum channel. 
\end{abstract}

\section{Introduction}
Quantum networks play an important role in scaling up quantum technologies by facilitating the exchange of quantum states and entanglement between otherwise separated quantum processing or memory units~\cite{Cirac1997,Kimble2008,Northup2014}. For such applications, residual thermal excitations in the connecting quantum channels are usually considered detrimental, since they spoil the transmitted quantum state or lead to false counts in heralded entanglement distribution schemes. Therefore, with the exception of purely linear quantum networks~\cite{Xiang2017,Vermersch2017,Xiang2023}, thermal noise at the relevant communication frequencies must be avoided. While fulfilling this requirement is usually not a problem in the optical regime~\cite{Reiserer2015}, it imposes considerable cooling requirements for superconducting~\cite{Magnard2020,Yam2025,Qiu2025, Mollenhauer2025} or phononic~\cite{Habraken2012,Gustafsson2014,Schuetz2015,Lemond2018,Bienfait2019,Dumur2021} quantum networks operating at much lower frequencies.

In this paper, we present a very different perspective on this problem and demonstrate that, rather than avoiding it, thermal noise can be harnessed as a useful resource for quantum communication applications. In the considered setup depicted in Fig.~\ref{Fig1:Setup}, two spatially separated qubits are coupled via a coherent quantum channel that is connected at its input to a purely thermal photon source. As a result, the qubits are driven by a field with an unknown and highly fluctuating amplitude and phase, and no coherence or entanglement is expected in such a scenario. Surprisingly, we find that when the bandwidth of the thermal source is sufficiently reduced—i.e., the reservoir becomes highly non-Markovian—the two qubits relax into a steady state with a nonvanishing amount of entanglement. This steady-state entanglement can be further enhanced in a systematic manner by decreasing the source’s bandwidth while increasing its temperature.
	
The emergence of such a thermally driven, delocalized entangled state is interesting both from a conceptual and a practical point of view. First of all, this effect clearly illustrates how the degree of non-Markovianity of a high-temperature thermal reservoir can strongly affect the qualitative features of a quantum system coupled to it. Secondly, while the current mechanism does not avoid the requirement of cooling the channel itself, it provides a completely passive scheme to distribute entanglement between distant qubits, which does not require any coherent control. Compared to similar effects predicted for specific thermally driven systems with local interactions~\cite{Brask2015,Khandelwal2025}, the current mechanism can generate close-to-maximally entangled states over arbitrary distances. This feature becomes very useful for generating a steady reservoir of distributed entangled states, which can then be further purified and used for state teleportation or other quantum communication protocols~\cite{Wootters1996,Zoller1999,Macchiavello1999,Cleland2022}. As two specific implementations, we theoretically analyze the generation of entangled states between two superconducting qubits connected by a cryogenic quantum link and between two SiV centers in a diamond phononic network. In both cases, we find that filtered noise from a room temperature resistor or from a locally heated mechanical mode is enough to generate highly entangled states.   

\section{A thermally driven quantum link}
\label{Sec:ThermalWaveguide}
\begin{figure}
	\centering	\includegraphics[width=\columnwidth]{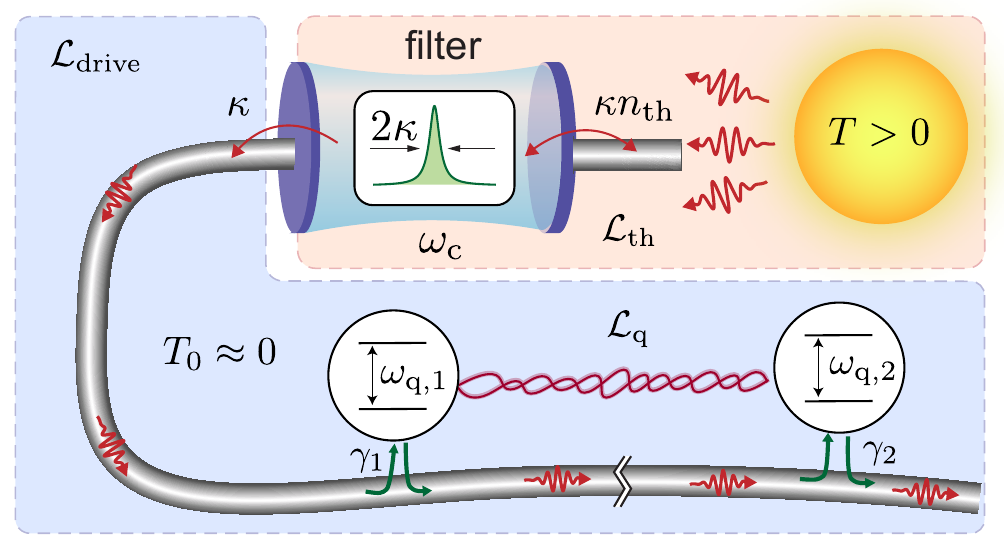}
	\caption{Sketch of a thermally driven quantum network, where two remote qubits are coupled via a unidirectional quantum channel. The qubits are located within the cold region of the network at a temperature $T_0$ far below their transition frequencies (blue shaded area). The input of the channel is connected to the output of a thermal photon source with a much higher temperature $T\gg T_0$ and a total bandwidth of $2\kappa$. This source can be realized by a two-sided filter cavity, which is coupled to the hot thermal radiation (right side) and the cold quantum channel (left side) with equal rates. See text for more details.
    }
	\label{Fig1:Setup}
\end{figure}

We consider a small quantum network as depicted in Fig.~\ref{Fig1:Setup}, where two spatially separated two-level systems (qubits) are coupled to a common bosonic quantum channel, for example, a photonic or phononic waveguide. The qubits with ground state $|0\rangle$ and excited state $|1\rangle$ have transition frequencies $\omega_{\mathrm{q},1}$ and $\omega_{\mathrm{q},2}$ and they decay into the waveguide with rates $\gamma_1$ and $\gamma_2$, respectively. To simplify the following analysis, we assume for now that this emission process is fully unidirectional, such that excitations in the waveguide propagate from qubit 1 to qubit 2, but not in the opposite direction. Chiral interactions of this type can be realized, for example, using  circulators~\cite{Sliwa15,Kerckoff15,Chapman17,Lecocq17,Masuda19,Wang21}, directional circuits~\cite{Guimond20,Gheeraert20,Kannan22,Joshi2023} or chiral waveguides~\cite{Lodahl17}. We emphasize, however, that such directional couplings are not a stringent requirement for the main effects discussed in this paper and in Sec.~\ref{Sec:Imp_Phon} below we will also consider alternative setups with bidirectional qubit-waveguide interactions.
The whole network is assumed to be cooled to a sufficiently low base temperature, $ T_0 \ll \hbar \omega_{\mathrm{q}, i}/k_B$ with $i=1,2$, such that in an undriven state, thermal excitations at the qubit frequencies can be neglected. To generate non-trivial qubit states, the input of the waveguide is then connected to a thermal source with a bandwidth of $2\kappa$. This source can be realized, for example, by a two-sided filter cavity with a central frequency $\omega_{\rm c} \sim \omega_{\mathrm{q}, i}$, which separates a region of high temperature $T\gg T_0$ from the low-temperature quantum channel. By assuming a symmetric cavity with the same loss rate $\kappa$ on both sides, the qubits are then driven by an average photon flux of $\Phi=\kappa n_{\rm th}/2$, where $n_{\rm th}=1/(e^{\hbar \omega_{\rm c}/(k_{B} T)}-1)$.

\subsection{Network master equation}
Under the assumption that the waveguide exhibits an approximately linear dispersion relation, its dynamics can be adiabatically eliminated to obtain a master equation for the density operator $\rho$, which describes the state of the qubits and the thermal source. For the current unidirectional setup, we apply the framework of cascaded quantum master equations~\cite{Carmichael93, Gardiner93,Zoller04}, whose further general properties can be found in Refs.~\cite{Lodahl17,Agusti2022,Zoller12,Pichler15}. We obtain 
\begin{equation}
	\label{Eq:ME_prelim}
	\dot{\rho}=( \mathcal{L}_{\rm th}+\mathcal{L}_{1}+\mathcal{L}_{2}+\mathcal{L}_{\rm casc})\rho,
\end{equation}
where the individual terms account for the dynamics of the thermal source, the dynamics of the individual qubits, and the waveguide-mediated interactions, respectively.  
	
By changing into a frame rotating with frequency $\omega_{\rm c}$, the filter cavity is described by the Liouville operator
\begin{equation}
	\label{Eq:ThermalLindblad}
	\begin{split}
		\mathcal{L}_{\rm th}\rho=\kappa (n_{\rm th}+1)\mathcal{D}[a]\rho+\kappa  n_{\rm th}\mathcal{D}[a^\dagger]\rho+\kappa  \mathcal{D}[a]\rho,
	\end{split}
\end{equation}
where $\mathcal{D}[O]\rho=O\rho O^\dagger-\{O^\dagger O, \rho\}/2$. Since we assume that the filter cavity is coupled symmetrically to the hot reservoir at temperature $T$ and to the cold waveguide at $T_0\approx 0$, the steady state generated by Eq.~\eqref{Eq:ThermalLindblad} is a thermal state with an average photon number of $\langle a^\dag a\rangle=n_{\rm th}/2$. In the same rotating frame, the dynamics of each individual qubit is given by $(\hbar=1)$
	\begin{equation}\label{eq:Li}
		\mathcal{L}_{i}\rho =-i[H_i,\rho]+\gamma_i \mathcal{D}[\sigma_i^-]\rho,
	\end{equation}
where $H_i= \Delta_i \sigma_i^z/2$ and $\Delta_i=\omega_{\mathrm{q},i}-\omega_{\mathrm{c}}$ is the detuning of the $i$-th qubit from the cavity resonance. Finally, the waveguide mediates a directional interaction between the cavity and the two qubits, which can be modeled in terms of a cascaded master equation~\cite{Carmichael93, Gardiner93, Zoller04} with Lindbladian\footnote{In deriving this result, we have neglected all propagation phases and retardation effects. However, in a purely unidirectional channel, Eq.~\eqref{Eq:CascadedInteraction} remains valid for arbitrary distances between the qubits and the cavity by introducing appropriately rotated basis states and time-shifted observables. See Ref.~\cite{Agusti2022} for a detailed discussion.} 
\begin{equation} 
	\label{Eq:CascadedInteraction}
	\begin{split} 
		\mathcal{L}_{\rm casc}\rho= &\sum_i \sqrt{\gamma_i\kappa} \left([a\rho,\sigma_i^+]+[\sigma_i^-,\rho a^\dagger]\right)\\
		&+\sqrt{\gamma_1\gamma_2}\left([\sigma_1^-\rho,\sigma_2^+]+[\sigma_2^-,\rho\sigma_1^+]\right).
	\end{split} 
\end{equation} 
Note that the full model in Eq.~\eqref{Eq:ME_prelim} describes an idealized scenario, where the qubits decay only into the waveguide and other losses and dephasing processes are neglected. The influence of those imperfections will be analyzed in Sec.~\ref{Sec:Implementations}. 
	
\subsection{Triplet-singlet basis} 
In our analysis below, we are primarily interested in the symmetric configuration $\gamma_1\simeq\gamma_2=\gamma$. In this case it is more convenient to regroup the cascaded interaction in Eq.~\eqref{Eq:CascadedInteraction} together with the dissipators in Eq.~\eqref{eq:Li} in terms of a single dissipation term with collective jump operator $S^-=(S^+)^\dag=\sigma_1^-+\sigma_2^-$ and a purely coherent term with Hamiltonian 
\begin{equation}
	\label{Eq:HCasc}
	H_{\rm casc}=i\frac{\gamma}{2}(\sigma_1^+\sigma_2^--\sigma_1^-\sigma_2^+).
\end{equation}
In this way, the full master equation in Eq.~\eqref{Eq:ME_prelim} can be rewritten in the form 
\begin{equation}
	\label{Eq:ME}
	\dot{\rho}=\left(\mathcal{L}_{\rm th}+\mathcal{L}_{\rm q}+\mathcal{L}_{\rm drive}\right)\rho.
\end{equation}
Here,
\begin{equation}\label{eq:Lq}
	\mathcal{L}_{\rm q}\rho=-i\left[H_{\rm q},\rho\right]+\gamma\mathcal{D}[S^-]\rho,
\end{equation}
with $H_{\rm q}=H_1+H_2+H_{\rm casc}$, describes the dissipative dynamics of the waveguide-coupled qubit system without the cavity, while 
\begin{equation}\label{eq:Ldrive}
	\mathcal{L}_{\rm drive}\rho=\sqrt{\gamma\kappa}\left([a\rho,S^+]+[S^-,\rho a^\dagger]\right),
\end{equation}
accounts for the effect of the cavity-filtered thermal radiation that is driving the qubits. We emphasize the asymmetric ordering of the cavity operators in this term, which ensures that the reduced cavity state remains unaffected by the qubit dynamics, i.e., ${\rm Tr}_{\rm q}\{\mathcal{L}_{\rm drive}\rho\}=0$, but not the other way around.

\begin{figure}
	\centering	\includegraphics[width=\columnwidth]{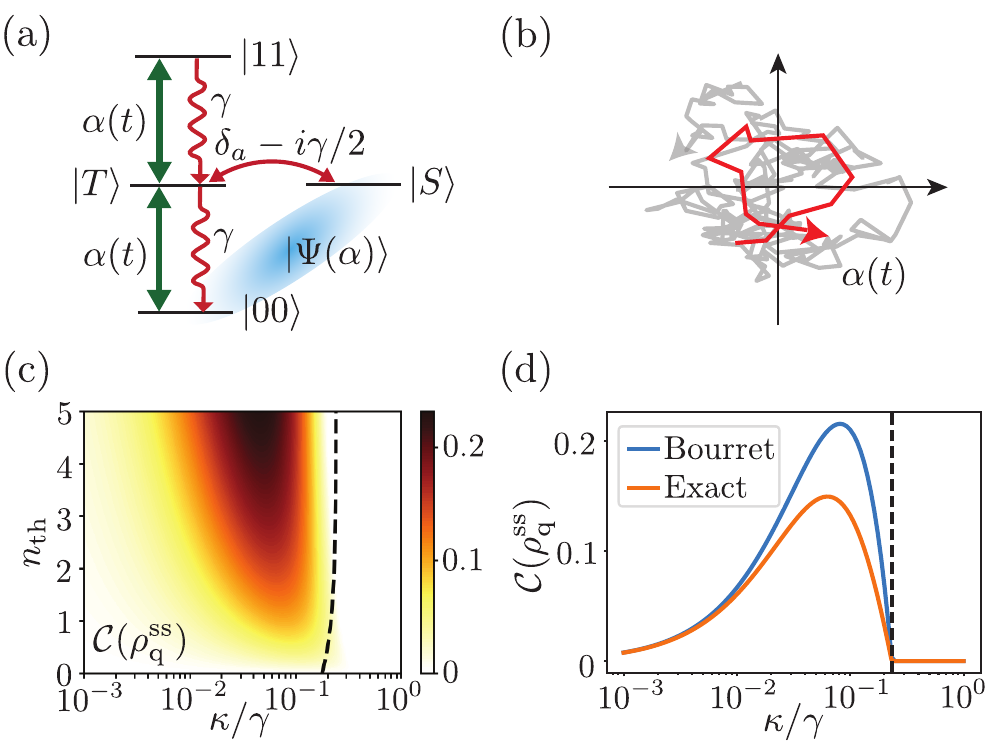}
	\caption{(a) Schematic representation of the two-qubit system in the triplet-singlet basis. (b) Illustration of the thermal driving field in phase space. The red and the grey trajectories represent the cases of a small bandwidth and of a large bandwidth, respectively.  (c) Steady-state entanglement of the reduced two-qubit state based on an exact numerical simulation of Eq.~\eqref{Eq:ME} and measured in terms of the concurrence $\mathcal{C}(\rho^{\rm ss}_{\rm q})$. (d) A slice of the data in (c) taken at a value of $n_{\rm th}=2$ is compared to the analytic result obtained within the Bourret approximation in Eq.~\eqref{Eq:ConcurrenceBourret}. In both plots, the dashed line corresponds to the maximal bandwidth for entanglement, $\kappa_{\rm max}/\gamma$, introduced above Eq.~\eqref{Eq:BandwidthCritical}.}
	\label{Fig2:Bandwidth}
\end{figure}
In this symmetric configuration, the combined states of two qubits can be best represented in a triplet-singlet basis, as depicted in Fig.~\ref{Fig2:Bandwidth}(a). In this basis, both the fully excited state $|11\rangle$ and the symmetric triplet state $|T\rangle =(|01\rangle + |10\rangle)/\sqrt{2}$ decay into waveguide. This follows from the second term in Eq.~\eqref{eq:Lq} and the action of the collective jump operator on those states, i.e., $S^-|11\rangle= \sqrt{2} |T\rangle$ and $S^-|T\rangle= \sqrt{2} |00\rangle$. In contrast, the anti-symmetric singlet state $|S\rangle =(|01\rangle - |10\rangle)/\sqrt{2}$ is a so-called dark state of the collective jump operator, i.e.,  $S^-|S\rangle =0$. Therefore, it is not affected by the dissipation term in Eq.~\eqref{eq:Lq}. However, the singlet and triplet states are coupled by the Hamiltonian $H_{\rm casc}$ and by any asymmetry in the detunings, $\delta_a=\Delta_1=-\Delta_2\neq 0$. As indicated in Fig.~\ref{Fig2:Bandwidth}(a), the corresponding coupling matrix element is 
\begin{equation}
\langle S |H_{\rm q}|T\rangle = \delta_{a} -i\frac{\gamma}{2}.
\end{equation}
Therefore, in this unidirectionally coupled system and in the absence of any driving fields, there is no complete destructive interference and the singlet state is not a dark state of the full dynamics described by Eq.~\eqref{eq:Lq}. Any initial population in this state is coherently transferred to the triplet state, from where it eventually decays to the ground state $|00\rangle$. The thermal source, which also drives the qubits symmetrically through the term $\mathcal{L}_{\rm drive}$ in Eq.~\eqref{eq:Ldrive}, induces transitions between the ground state $|00\rangle$ and $|T\rangle$ and between $|T\rangle$ and $|11\rangle$, respectively. Again, the singlet state is not directly affected by this process, since $S^-|S\rangle= S^+|S\rangle=0$.
	
\subsection{Stochastic dynamics in phase space}
\label{Sec:StochasticME}
	
While for moderate temperatures with $n_{\rm th} \sim O(10)$ the dynamics and steady states of Eq.~\eqref{Eq:ME} can be evaluated numerically in a straightforward manner, this is no longer possible for much larger thermal occupation numbers, where a representation of the master equation in terms of the usual number states for the cavity mode becomes very memory-inefficient. To treat this high-temperature limit, it is more convenient to switch to a phase-space representation, where the cavity is described by a distribution over coherent states $|\alpha\rangle$ with complex amplitudes $\alpha \in \mathbbm{C}$. In particular, for the case of a purely thermal source, we can express the full system density operator as~\cite{Ritsch1988,Zoller04}
\begin{equation}
	\label{Eq:FPrho}
	\rho(t)= \int {\rm d}^2 \alpha \,   |\alpha\rangle \langle \alpha |  \otimes \mu(\alpha,t),
\end{equation} 
where $\mu(\alpha,t)$ is an operator in the two-qubit subspace. By taking partial traces over both sides of Eq.~\eqref{Eq:FPrho}, we can identify the integral 
\begin{equation}
\rho_{\rm q} (t)= \int {\rm d}^2 \alpha \, \mu(\alpha,t),
\end{equation}
with the reduced state of the two qubits and 
\begin{equation}
P(\alpha,t) = {\rm Tr}_{\rm q}\{\mu(\alpha,t)\},
\end{equation}
with the Glauber-Sudarshan P-function of the cavity mode.
 
By inserting the ansatz in Eq.~\eqref{Eq:FPrho} into the master equation in Eq.~\eqref{Eq:ME} and using the usual mapping between density operators and phase-space distributions~\cite{Zoller04}, we find that the qubit operator $\mu(\alpha,t)$ obeys the following equation of motion
	\begin{equation}
		\label{Eq:FokkerPlankME}
		\left(\frac{d}{dt}+\Lambda \right)\mu(\alpha,t)=\left( \mathcal{L}_{\rm q} + \alpha \mathcal{L}_+ + \alpha^* \mathcal{L}_-\right)\mu(\alpha,t),
	\end{equation}
where $\mathcal{L}_{\pm}\rho=\pm\sqrt{\gamma\kappa} [\rho,S^\pm]$. Here we have introduced the differential operator~\cite{Risken1996}
\begin{equation}\label{eq:Ldiffop}
	\Lambda= - \kappa\left( \frac{\partial }{\partial \alpha}  \alpha +  \frac{\partial }{\partial \alpha^*} \alpha^*+ n_{\rm th} \frac{\partial^2}{\partial \alpha \partial \alpha^*} \right),
\end{equation}
which describes the evolution of the P-function of the bare thermal cavity mode. 

Equation~\eqref{Eq:FokkerPlankME} determines the evolution of an operator-valued distribution in phase space. This distribution can be sampled in terms of a set of random trajectories for the amplitudes $\alpha(t)$ and the corresponding evolution of $\mu(t)\equiv \mu(\alpha(t),t)$~\cite{Ritsch1988}. In our unidirectional setting, the thermal cavity is not affected by the qubits and therefore the amplitudes $\alpha(t)$ obey the stochastic (Itô) differential equation 
	\begin{equation}
		\label{Eq:StochasticOU}
		{\rm d}\alpha(t) =-\kappa \alpha(t)  {\rm d}t +\sqrt{\kappa n_{\rm th}} {\rm d}W(t),
	\end{equation}
	where ${\rm d}W(t)$ is a complex Wiener increment. In turn, for a given $\alpha(t)$, the corresponding qubit density operator evolves as
	\begin{equation}
		\label{Eq:StochasticME}
		\dot{\mu}(t)=-i\left[ H_\alpha(t) ,\mu(t)\right]+\gamma\mathcal{D}[S^-]\mu(t).
	\end{equation}
Here we have introduced the time-dependent Hamiltonian 
	\begin{equation}
		H_\alpha(t)= H_{\rm q} - i \sqrt{\kappa\gamma}\left[ \alpha(t) S^+- \alpha^{*}(t) S^-\right],
	\end{equation} 
describing the qubits driven by a fluctuating classical field [as depicted in Fig.~\ref{Fig2:Bandwidth}(b)]. 
The actual qubit density operator, $\rho_{\rm q}(t)=\langle \mu(\alpha(t),t) \rangle_{\rm st}$, can then be obtained from a combined numerical simulation of Eq.~\eqref{Eq:StochasticOU} and Eq.~\eqref{Eq:StochasticME} and after averaging over sufficiently many stochastic trajectories, denoted by $\langle \cdot \rangle_{\rm st}$. This approach also allows us to investigate the evolution of individual trajectories, which more accurately represent the system during a single experimental run. 
	
\section{Steady State Entanglement}\label{Sec:SteadyStateEntanglement}
In the following, we are interested in the stationary two-qubit state $\rho_{\rm q}^{\rm ss}={\rm Tr}_{\rm th} \{ \rho(t\rightarrow \infty)\}$. For this state, we quantify the amount of entanglement by the concurrence $\mathcal{C}(\rho_{\rm q})$~\cite{Hill97,Horodecki09}, which vanishes for any non-entangled state and reaches a maximal value of  $\mathcal{C}(\rho_{\rm q})=1$ for any of the Bell states. In Fig.~\ref{Fig2:Bandwidth}(c) we use exact numerical simulations of Eq.~\eqref{Eq:ME} to evaluate the steady-state entanglement, $\mathcal{C}(\rho_{\rm q}^{\rm ss})$, for different ratios of $\kappa/\gamma$ and moderate thermal occupation numbers $n_{\rm th}$. We clearly see the absence of entanglement in the Markovian regime, $\kappa\gtrsim \gamma$, while a considerable amount of entanglement can be observed for a non-Markovian photon source with $\kappa \ll \gamma$. Surprisingly, the maximal amount of entanglement increases with increasing temperature of the reservoir. 

\subsection{Markov limit}
In the derivation of the master equation in Eq.~\eqref{Eq:ME_prelim}, we have assumed that the waveguide connecting the source and the qubits is sufficiently broadband, and we have used a Markov approximation to eliminate its dynamics. However, in this approach, we still retain the exact dynamics of the source, which evolves on a timescale set by $\kappa^{-1}$. By making the additional assumption that $\kappa \gg\gamma$, we can also treat the thermal cavity as an effective Markovian reservoir and derive a master equation for the reduced two-qubit state $\rho_{\rm q}(t)={\rm Tr}_{\rm th}\{\rho(t)\}$~\cite{gonzalez-ballestero2024tutorial,carmichael2007statistical2}. 
This master equation reads 
\begin{equation}
	\label{Eq:MarkovianME}
	\begin{split}
		\dot{\rho}_{\rm q}=&-i[H_{\rm q},\rho_{\rm q}]+\gamma(n_{\rm th}+1)\mathcal{D}[S^-]\rho_{\rm q}\\
        &+ \gamma n_{\rm th}\mathcal{D}[S^+]\rho_{\rm q},
		\end{split}
\end{equation}
and describes the case of two qubits coupled to a (unidirectional) thermal environment. 

Starting from Eq.~\eqref{Eq:MarkovianME}, we can derive a closed set of equations for the steady-state matrix elements $\rho^{\rm ss}_{ij,kl}=\langle i,j|\rho_{\rm q}^{\rm ss}|k,l\rangle$. We find that the only non-zero matrix elements are
	\begin{equation}
		\label{Eq:ME_Markov}
		\begin{split}
			&\rho^{\rm ss}_{0}=(n_{\rm th}+1)^2/(1+2n_{\rm th})^2,\\
			\rho^{\rm ss}_{T}=&\rho^{\rm ss}_{S}=n_{\rm th}(n_{\rm th}+1)/(1+2n_{\rm th})^2,\\
			&\rho^{\rm ss}_{1}=n^2_{\rm th}/(1+2n_{\rm th})^2,
		\end{split}
	\end{equation}
where $\rho_{S/T}=\langle S/T|\rho_{\rm q}| S/T\rangle$ denote the singlet and triplet populations and we also introduced the short-hand notation $\rho_{i}\equiv \rho_{ii,ii}$ for the populations. These matrix elements correspond to the fully separable state $\rho_{\rm q}^{\rm ss}=\rho_{\rm th}\otimes \rho_{\rm th}$, with each of the qubits being in a thermal equilibrium state. This explains the absence of entanglement in the Markovian regime for any $n_{\rm th}$ and shows that even classical correlations, which could be mediated by the waveguide, are absent in this limit.

\subsection{Quasistatic limit}\label{Sec:Quasistatic}
To explain the existence of a nonvanishing amount of steady-state entanglement, we focus on the opposite limit,  where the evolution of the cavity mode is very slow compared to the relaxation time of the qubits, $\gamma^{-1}$. In this quasistatic limit, it is more convenient to work with the phase-space representation introduced in Sec.~\ref{Sec:StochasticME}. In this picture,  the qubits are driven by a classical field $\alpha(t)$, which fluctuates on a timescale set by $\kappa^{-1}$. Therefore, in the limit $\kappa\rightarrow 0$ (while keeping the photon flux $\Phi=\kappa n_{\rm th}/2$ finite) we can assume that this field is approximately constant and evaluate the steady state of the qubits for a fixed amplitude $\alpha(t)\approx \alpha_0$. For this case and assuming $\Delta_i=0$, it has been previously shown that the qubits relax into a pure steady state, $\rho_{\rm q}^{\rm ss}(\alpha_0)= |\Psi(\alpha_0)\rangle\langle \Psi(\alpha_0)|$~\cite{Zoller12,Pichler15}, where
	\begin{equation}\label{Eq:StaticPureState}
		|\Psi(\alpha_0)\rangle = \frac{\sqrt{\gamma}|00\rangle +  2\sqrt{2\kappa}\alpha_0 |S\rangle}{\sqrt{\gamma+8\kappa |\alpha_0|^2}}.  
	\end{equation}
Since this superposition involves only the states $|00\rangle$ and $|S\rangle$, it is a dark state of the collective jump operator,  i.e. $S^-|\Psi(\alpha_0)\rangle=0$. At the same time,  transitions to the triplet state $|T\rangle$ are cancelled by destructive interference between the driving term and the coupling induced by $H_{\rm casc}$ [see the level diagram in Fig.~\ref{Fig2:Bandwidth}(a)]. Therefore, the state is also a dark state of the Hamiltonian evolution, i.e., $H_{\alpha_0}|\Psi(\alpha_0)\rangle=0$. 
    
To obtain the actual two-qubit steady state, the result from above must be averaged over the thermal distribution of complex amplitudes $\alpha_0$, i.e., 
\begin{equation}
	\label{Eq:OUAverage}
	\rho^{\rm ss}_{\rm q}=\frac{2}{\pi n_{\rm th}}\int {\rm d}^2\alpha_0 |\Psi(\alpha_0)\rangle\langle\Psi(\alpha_0)| e^{-\frac{2|\alpha_0|^2}{n_{\rm th}}}.
\end{equation}
By carrying out this integral, we obtain
	\begin{equation}
		\label{Eq:StaticAvgState}
		\rho^{\rm ss}_{\rm q}=\Upsilon(\Phi/\gamma) |00\rangle\langle 00|  +\left[1-\Upsilon(\Phi/\gamma)\right] |S\rangle\langle S|,
	\end{equation}
where $\Upsilon(x)=e^{1/(8x)}\Gamma[0,1/(8x)]/(8x)$ and $\Gamma[0,x]$ is the upper incomplete gamma function. 
	
In the absence of any coherences between the states $|00\rangle$, $|S\rangle$, $|T\rangle$ and $|11\rangle$ and assuming that the singlet population $\rho_{S}$ is larger than both $\rho_{T}$ and $\sqrt{\rho_{0}\rho_{1}}$, the expression for the concurrence of the two-qubit state simplifies to  
	\begin{equation}
		\label{Eq:Concurrence}
		\mathcal{C}(\rho_{\rm q})={\rm max}\{0,\rho_S-\rho_T-2\sqrt{\rho_{0}\rho_{1}}\}.
	\end{equation}
 Therefore, for the steady state given in Eq.~\eqref{Eq:StaticAvgState}, we obtain
	\begin{equation}
		\label{Eq:AnalyticConcurrence}
		\mathcal{C}(\rho_{\rm q}^{\rm ss})=1-\Upsilon(\Phi/\gamma),
	\end{equation}
which grows as $\mathcal{C}(\rho_{\rm q}^{\rm ss})\simeq 8 \Phi/\gamma$ for a weak photon flux and it reaches a maximal value of $\mathcal{C}(\rho_{\rm q}^{\rm ss})\simeq 1-\gamma/(8\Phi)\rightarrow 1$ in the high-temperature limit.

\subsection{Discussion: coherent vs. thermal fields}
The numerical and analytical results presented above illustrate very clearly the vastly different roles that thermal reservoirs can play in the respective Markovian and non-Markovian limits. The analytic findings obtained in the quasistatic regime also show that the emergence of entanglement in our setup relies on the combination of several ingredients. First of all, the thermal source drives both qubits with the same phase, such that the singlet state $|S\rangle$, which is also stable against losses, decouples from the noisy field. This combination is necessary to permit the formation of the dark superposition state in Eq.~\eqref{Eq:StaticPureState} for the fixed amplitude $\alpha_0$.
Second, the actual field amplitude $\alpha(t)$ changes sufficiently slowly in time such that the instantaneous state of the systems follows this dark superposition state adiabatically. Fast variations of $\alpha(t)$ would cause non-adiabatic transitions out of this state and result in a nonvanishing population of the triplet and the doubly excited state.

Equally important, however, is the third property, namely that the generated entanglement does not rely on a fixed phase relation between $|00\rangle$ and $|S\rangle$. This fact can be seen from the full density operator in Eq.~\eqref{Eq:StaticAvgState}, which is the state after averaging over the unknown phase and amplitude of the thermal field. The entanglement of this state depends on the population of the singlet state only and is, to a good approximation, insensitive to the phase and the precise value of the amplitude of $\alpha_0$. For comparison, an instantaneous state of the form $\sim(|01\rangle+\alpha_0|10\rangle)$ would not have this property and lose its entanglement, once averaged over a thermal distribution of amplitudes $\alpha_0$. 

\begin{figure}
		\centering	\includegraphics[width=\columnwidth]{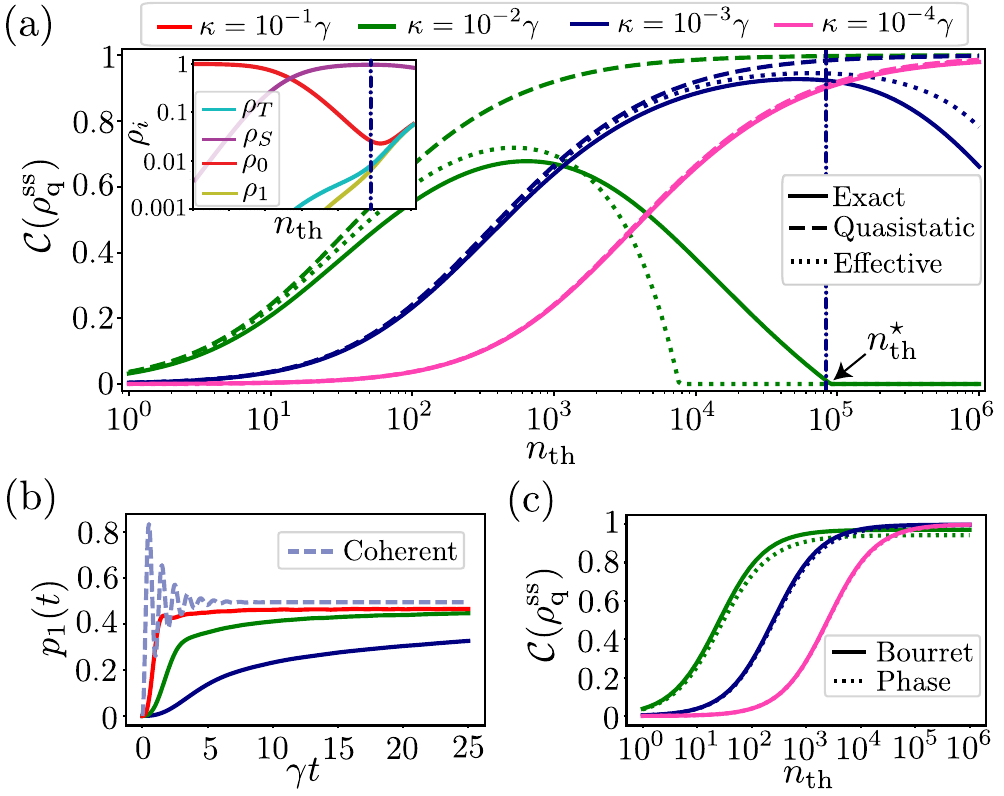}
		\caption{(a) Plot of the exact value of the steady-state concurrence $\mathcal{C}(\rho_{\rm q}^{\rm ss})$ (solid lines) for different ratios $\kappa/\gamma$ and as a function of the thermal occupation number, $n_{\rm th}$. The dashed lines show the approximate result derived in Eq.~\eqref{Eq:AnalyticConcurrence} for the quasi-static limit and the dotted line represents the predictions obtained in Eq.~\eqref{Eq:ConcurrenceEffective} using a continued fraction expansion. The inset shows the steady state populations of the two-qubit state for $\kappa/\gamma=10^{-3}$. Also for this bandwidth, the expression for $n_{\rm th}^{\star}\equiv n^\star_{\rm th}(\kappa=10^{-3}\gamma)$ given in Eq.~\eqref{Eq:ThermalCritic} is indicated by the vertical dash-dotted line. (b) Evolution of the excited state population $p_1(t)$ of a single qubit driven by a thermal source of varying bandwidth, assuming that the photon flux $\Phi=10\gamma$ is held fixed. For comparison, the dashed line indicates the evolution for a coherently driven qubit with Rabi frequency $\Omega_R=2\sqrt{\gamma \Phi}$. (c) Steady-state concurrence $\mathcal{C}(\rho_{\rm q}^{\rm ss})$ as predicted by the Bourret approximation described in Sec.~\ref{Sec:BourretApproximation} (solid lines) and by the phase-fluctuation model developed in Sec.~\ref{Sec:Phase} (dotted lines). In all plots, $\Delta_i=0$.}
		\label{Fig3:Populations}
	\end{figure}
    
To further elaborate on this point, let us emphasize that even a highly filtered thermal source is not equivalent to a coherent driving field. While the output of the filter cavity exhibits first-order coherence,
\begin{equation}
g^{(1)}(\tau)= \frac{\langle a^\dagger (\tau) a(0)\rangle}{\langle a^\dag a\rangle} = e^{-\kappa\tau},
\end{equation}
which extends over times much longer than the relaxation time of the qubits, its second-order coherence, 
\begin{equation}
g^{(2)}(\tau)= \frac{\langle a^\dagger(0) a^\dagger (\tau) a(\tau) a(0)\rangle}{\langle a^\dag a\rangle^2} = 1+ e^{-2\kappa \tau},
\end{equation}
corresponds to that of a thermal state over the same timescale. To illustrate this difference, in Fig.~\ref{Fig3:Populations}(b) we set $\gamma_2=0$ and simulate the driving of a single qubit by a thermal source with a fixed photon flux $\Phi=10\gamma$, but varying $\kappa$. We see that instead of Rabi oscillations with frequency $\Omega_R=2\sqrt{\gamma \Phi}$, as expected for a coherent driving field with fixed amplitude $\alpha_0=\sqrt{n_{\rm th}/2}$, we observe a monotonic increase of the population, due to an averaging over a thermal distribution of amplitudes. Therefore, the dynamics of a quantum system driven by a narrow-bandwidth thermal source differs from that of a coherently driven system. For our entanglement mechanism, it is essential that the specific dark state in Eq.~\eqref{Eq:StaticAvgState} is not considerably affected by this difference, because it is highly insensitive against thermal uncertainties in both the phase and the amplitude of the driving field.

Equation~\eqref{Eq:AnalyticConcurrence} predicts that the entanglement increases with the temperature of the reservoir and saturates at very high values of $\Phi\gg \gamma$. However, this result has been derived in the static limit $\kappa\rightarrow 0$, where in a strict sense also the photon flux $\Phi$ vanishes. Therefore, to prove the existence of a highly-entangled steady state, we must go beyond the analysis of Sec.~\ref{Sec:Quasistatic}. In particular, non-adiabatic transitions into the triplet state $|T\rangle$ or the fully excited state $|11\rangle$ must be taken into account, which are expected to play a more and more important role as the temperature increases. 
			
\section{Entanglement in a non-Markovian, high-temperature reservoir}
\label{Sec:BeyondStatic}
	
To determine the steady state of the two qubits in the most interesting parameter regime, we must investigate the dynamics and steady states of the network at nonvanishing, but very small ratios $\kappa/\gamma$ and for very high thermal occupation numbers $n_{\rm th}$. In Sec.~\ref{Sec:ContinuedFraction} below, we describe a method based on a matrix continued fraction expansion, which allows us to numerically evaluate the exact steady state 
$\rho_{\rm q}^{\rm ss}$ for arbitrary temperatures. The main results from these simulations are summarized in Fig.~\ref{Fig3:Populations}(a), which shows the behavior of the concurrence and the steady-state populations as a function of $n_{\rm th}$ and for different ratios $\kappa/\gamma$. While for small and moderate temperatures the predictions of the quasistatic limit given in Sec.~\ref{Sec:Quasistatic} are very accurately reproduced, there exists a characteristic thermal occupation number $n^\star_{\rm th} \equiv n^\star_{\rm th}(\kappa/\gamma)$, beyond which the entanglement degrades again. We attribute this degradation to non-adiabatic transitions out of the quasiadiabatic dark state, which become more and more relevant beyond this point. In the remainder of this section, we present additional analytic approximations to obtain further insights into the behavior of the system across a wider parameter range and to predict the cross-over thermal occupation number $n^\star_{\mathrm{th}}$.   

\subsection{Decorrelation approximation}\label{Sec:BourretApproximation}
In a first step, to interpolate between the Markovian and the quasistatic regime, we use a Bourret approximation to derive a closed, but time-nonlocal equation for the qubit state. As detailed in Appendix~\ref{Sec:Bourret}, under this approximation we obtain a time-nonlocal equation for the qubit state,
\begin{equation}\label{Eq:WeakME}
\begin{split}
			&\dot{\rho}_{\rm q}(t)  \simeq \mathcal{L}_{\rm q}\rho_{\rm q}(t)\\
			+&\int_{0}^t \mathrm{d}t' \,  \langle \alpha^*(t)\alpha(t')\rangle_{\rm st}  \,  \mathcal{L}_+ e^{\mathcal{L}_{\rm q}(t-t')} \mathcal{L}_- \rho_{\rm q}(t')+ {\rm H.c.},
\end{split}
\end{equation}
where $\langle \alpha^*(t)\alpha(t')\rangle_{\rm st} = n_{\rm th} e^{-\kappa|t-t'|}/2$ for our thermal noise source. Note that the derivation of Eq.~\eqref{Eq:WeakME} relies on a factorization of the expectation value $\langle \alpha^*(t)\alpha(t')\mu(t')\rangle_{\rm st}\simeq \langle \alpha^*(t)\alpha(t')\rangle_{\rm st} \langle\mu(t')\rangle_{\rm st}$, where $\mu(t)\equiv \mu(\alpha(t),t)$ is the qubit operator introduced in Eq.~\eqref{Eq:FPrho}. In a strict sense, this approximation is only valid in the regime of very weak thermal excitations. Nevertheless, we can use Eq.~\eqref{Eq:WeakME} to obtain an approximate two-qubit state at arbitrary temperatures.

Under the Bourret approximation, the steady state of the qubits can be obtained by taking the Laplace transform of Eq.~\eqref{Eq:WeakME}. As shown in Appendix~\ref{Sec:Bourret}, the only non-vanishing matrix elements are the populations $\rho^{\rm ss}_{0}$, $\rho^{\rm ss}_S$, $\rho^{\rm ss}_T$ and $\rho^{\rm ss}_{1}$, which, up to the lowest relevant order in $\kappa$, are given by
\begin{align}
		&\rho^{\rm ss}_{1}\simeq\frac{64\kappa \Phi^2}{\gamma(\gamma+8\Phi)(9\gamma+8\Phi)},\\
		&\rho^{\rm ss}_T\simeq\frac{8\kappa \Phi}{\gamma(\gamma+8\Phi)},\\
		&\rho^{\rm ss}_S\simeq\frac{8\Phi}{\gamma+8\Phi}-\frac{8\kappa \Phi(27\gamma^2+112\gamma \Phi+192\Phi^2)}{\gamma(\gamma+8\Phi)^2(9\gamma+8\Phi)},\\
        &\rho^{\rm ss}_{0}\simeq\frac{\gamma}{\gamma+8\Phi}.
\end{align}
From these expressions, we can use Eq.~\eqref{Eq:Concurrence} to evaluate the concurrence, 
\begin{equation}\label{Eq:ConcurrenceBourret}
    \begin{split}
    \mathcal{C}(\rho_{\rm q}^{\rm ss})\simeq&\frac{8\Phi}{\gamma+8\Phi}-\frac{16\Phi \sqrt{\kappa}}{(\gamma+8\Phi)\sqrt{9\gamma+8\Phi}}\\
    &-\frac{32 \kappa \Phi(3\gamma+8\Phi )^2}{\gamma(\gamma+8\Phi)^2(9\gamma+8\Phi)}.
    \end{split}
\end{equation}
In Fig.~\ref{Fig2:Bandwidth}(d), this approximate form of the concurrence is compared with the exact result for a small value of $n_{\rm th}=2$. It can also be used to evaluate the maximal bandwidth $\kappa_{\rm max}$, up to which steady-state entanglement can be observed. 
This boundary between the Markovian and the non-Markovian regime is indicated by the dashed line in Fig.~\ref{Fig2:Bandwidth}(c) and (d). It falls within the bounds 
\begin{equation}\label{Eq:BandwidthCritical}
0.18 \lesssim  \frac{\kappa_{\rm max}}{\gamma} < 0.25,
\end{equation}
with only a weak dependence on $n_{\rm th}$. For a bandwidth below this boundary the concurrence increases with increasing photon flux $\Phi$ and saturates at a value of 
\begin{equation}\label{Eq:Bourret_Concurrence_Saturation}
    \mathcal{C}(\rho_{\rm q}^{\rm ss})\approx 1-\frac{4\kappa}{\gamma}.
\end{equation}
This value is reduced compared to the predictions of the purely static limit, but only by a small amount. The concurrence given in Eq.~\eqref{Eq:ConcurrenceBourret} also does not exhibit a maximum at finite $n_{\rm th}^\star$, as observed in exact numerical simulations [see also Fig.~\ref{Fig3:Populations}(c)]. Therefore, while the Bourret approximation captures very well the transition between the Markovian and the non-Markovian regime at low and moderate temperatures, it fails to predict the behavior of the system in the strongly driven regime, $\Phi/\gamma\gtrsim 1$, where the decorrelation approximation breaks down. Note that also other types of second-order approximations, such as the time-local master equations discussed in Ref.~\cite{Groszkowski2023}, fail to capture the qualitative behavior of our system in the non-Markovian, high-temperature regime.

\subsection{Phase vs. amplitude fluctuations}\label{Sec:Phase}
For a slowly varying $\alpha(t)$, the qubits will relax into the superposition state given in Eq.~\eqref{Eq:StaticAvgState}, but with time-dependent coefficients, $\alpha_0\rightarrow \alpha(t)$. Slow changes of either the phase or the amplitude of $\alpha(t)$ can then induce non-adiabatic transitions into the triplet state and spoil the entanglement. To understand which type of fluctuations are most detrimental, we express the field amplitude in polar coordinates as
	\begin{equation}
		\alpha(t) = r(t) e^{i\theta(t)}.
	\end{equation} 
Starting from the stochastic equation for $\mathrm{d}\alpha(t)$ in Eq.~\eqref{Eq:StochasticOU}, we can apply the rules of Ito calculus to derive the corresponding coupled stochastic updates for the radius $r(t)$ and the phase $\theta(t)$. We obtain~\cite{Zoller2014}
	\begin{flalign}
		& {\rm d}r (t)= \left(-\kappa r(t)   + \frac{\kappa n_{\rm th}}{4 r(t)} \right){\rm d}t +\sqrt{\frac{\kappa n_{\rm th}}{2}}{\rm d}W_{r}(t), \\
		& {\rm d}\theta (t)=\sqrt{\frac{\kappa n_{\rm th}}{2}}\frac{{\rm d}W_{\theta}(t)}{r(t)},\label{Eq:Phase}
	\end{flalign}
	with two independent and real-valued Wiener increments, ${\rm d}W_r(t)$ and ${\rm d}W_\theta(t)$. 
	
In contrast to the complex-valued Ornstein-Uhlenbeck process in Eq.~\eqref{Eq:StochasticOU}, the stochastic differential equations in polar coordinates cannot be solved exactly, but they allow us to investigate the independent role of phase fluctuations by fixing the radial component to $r(t)=r_0$. In this case, the phase obeys a pure diffusion process ${\rm d}\theta= \sqrt{\kappa n_{\rm th}/(2r_0^2)}{\rm d}W_{\theta}(t)$. To solve the corresponding stochastic master equation for the qubits, we start from Eq.~\eqref{Eq:StochasticME} and change into a co-rotating frame~\cite{Agarwal1978}, $\tilde{\mu}_{\rm q}(\theta(t),t)=e^{-i\theta(t) S^z/2}\mu_{\rm q}(\theta(t),t)e^{i\theta(t) S^z/2}$, where $S^z=\sigma^z_1+\sigma^z_2$. In this new basis, the resulting stochastic master equation for $\tilde\mu_{\rm q}(t)\equiv \tilde{\mu}_{\rm q}(\theta(t),t)$ reads
	\begin{equation}\label{Eq:PhaseStochasticME}
	\begin{split}
		\dot{\tilde{\mu}}_{\rm q}(t)=\, &\left[\mathcal{L}_0+r_0\left(\mathcal{L}_++\mathcal{L}_-\right)\right]\tilde{\mu}_{\rm q}(t)-i\frac{\dot{\theta}}{2}[S^z,\tilde{\mu}_{\rm q}(t)].
	\end{split}
	\end{equation}
This equation must be interpreted as a Stratonovich stochastic differential equation, where $\dot \theta$ and $\tilde{\mu}_{\rm q}(\theta(t),t)$ are not independent. To proceed, we convert Eq.~\eqref{Eq:PhaseStochasticME}  to Itô form, after which we can take the average over the independent Wiener increments. We finally end up with a deterministic master equation for $\tilde{\rho}_{\rm q}(t)=\langle \tilde{\mu}_{\rm q}(\theta(t),t)\rangle_{\mathrm{st}}$, which is given by
\begin{equation}\label{Eq:PhaseAvgME}
		\dot{\tilde{\rho}}_{\rm q}=\left[\mathcal{L}_0+r_0\left(\mathcal{L}_++\mathcal{L}_-\right)\right]\tilde{\rho}_{\rm q}+\frac{\kappa n_{\rm th}}{8 r_0^2}\mathcal{D}[S^z]\tilde{\rho}_{\rm q}.
\end{equation}
 Note that while $\tilde{\rho}_{\rm q}(t)$ is the state in a rotated basis, the unitary phase rotation introduced above does not affect the diagonal elements and we can identify $\rho^{\rm ss}_{ij,ij}=\langle i,j|\tilde \rho_{\rm q}^{\rm ss}|i,j\rangle$.

Compared to a classical driving field with fixed amplitude $r_0$, the master equation in Eq.~\eqref{Eq:PhaseAvgME} contains an additional collective dephasing term. This term affects the destructive interference responsible for the existence of the dark state in Eq.~\eqref{Eq:StaticAvgState} and facilitates excitations of states $|T\rangle$ and $|11\rangle$. However, by replacing $r_0$ with the mean value $r_0=\sqrt{n_{\rm th}/2}$, we find that this dephasing term only scales with the bandwidth $\kappa$ and not with the thermal occupation number. In Fig.~\ref{Fig3:Populations}(c) we plot the predictions from Eq.~\eqref{Eq:PhaseAvgME} for different parameters and from a comparison with simulations of the full system, we find that Eq.~\eqref{Eq:PhaseAvgME} in general underestimates the effect of temporal fluctuations and predicts a steady-state concurrence that increases monotonically with $n_{\rm th}$. Therefore, phase fluctuations do not explain the observed behavior in the high-temperature regime. We conclude that the main correction to the quasiadiabatic limit arises from fluctuations of the absolute value of the field amplitude, $r(t)=|\alpha(t)|$. Unfortunately, a similar closed equation as above cannot be obtained for a fluctuating $r(t)$.

\subsection{Strong thermal driving}\label{Sec:ContinuedFraction}
Our discussion above shows that neither a decorrelation approximation nor a simple phase-diffusion model is able to correctly predict the qualitative behavior of our system in the most relevant regime where the qubits are driven by slow but large-amplitude thermal fluctuations. Therefore, in Appendix~\ref{AppSec:ContinuedFraction} we outline a more general approach to address this limit, following similar ideas as developed in Ref.~\cite{zoller1979resonant,zoller1979stark,cirac1991thermal} for a single two-level system. 

The starting point of this analysis is the phase-space representation of the density operator in Eq.~\eqref{Eq:FPrho}, which obeys the equation of motion given in Eq.~\eqref{Eq:FokkerPlankME}. A solution to this equation can be obtained by expanding $\mu(\alpha,t)$ in terms of  a complete biorthogonal set of eigenfunctions $P_{n,m}$ and $\phi_{n,m}$ of the differential operator $\Lambda$, defined as~\cite{zoller1979resonant,zoller1979stark}
\begin{flalign}
\Lambda P_{n,m}=\lambda_{n,m}P_{n,m}, \qquad \Lambda^\dagger \phi_{n,m}=\lambda^*_{n,m}\phi_{n,m}.
	\end{flalign}
For the complex-valued Ornstein-Uhlenbeck process, the eigenvalues are $\lambda_{n,m}=\kappa(2n+|m|)$, with $n=0,1,2,\dots$ and $m=0,\pm1,\pm2,\dots$,  and the corresponding eigenmodes are related by $P_{n,m}(\alpha)=P_{\rm ss}(\alpha)\phi_{n,m}(\alpha)$. Here 
$P_{\rm ss}(\alpha)\equiv P_{0,0}(\alpha)=2/(\pi n_{\rm th})e^{-2|\alpha|^2/n_{\rm th}}$ is the steady-state distribution and 
\begin{equation}
\begin{split}
	\phi_{n,m}(\alpha)=&\sqrt{\frac{n!}{(n+|m|)!}\frac{|\alpha|^{2|m|}}{(n_{\rm th}/2)^{|m|}}\left(\frac{\alpha^*}{\alpha}\right)^{m} }\\
    &\times L_n^{|m|}\left[\frac{2|\alpha|^2}{n_{\rm th}}\right],
\end{split}
\end{equation}
where $L_n^m[x]$ are generalized Laguerre polynomials. By using the mode expansion
\begin{equation}\label{eq:muExpansion} 
\mu(\alpha, t) = \sum_{n,m} P_{n,m}(\alpha) \mu^{n,m}(t), 
\end{equation} 	
Eq.~\eqref{Eq:FokkerPlankME} can be converted into a set of coupled equations for the operators $\mu^{n,m}$, and since  $\phi_{0,0}(\alpha)=1$, we can identify the component 
\begin{equation}
\mu^{0,0}(t)= \int \mathrm{d}^2 \alpha \, \mu(\alpha,t) = \rho_{\rm q}(t)
\end{equation} 
with the reduced two-qubit state of interest.  As explained in more detail in Appendix~\ref{AppSec:ContinuedFraction}, the resulting recurrence relations for the elements $\mu^{n,m}$ in the steady state can then be solved for arbitrary values of $n_{\rm th}$ by numerically evaluating a matrix continued fraction.

In Appendix~\ref{AppSec:ContinuedFraction}, we also outline several further approximations to simplify the resulting set of equations and obtain analytical expressions for $\rho_{\rm q}^{\rm ss}$. This involves, first of all, a three-level approximation, where transitions to the state $|11\rangle$ are neglected. From numerical benchmarks, we find that this is still a good approximation for $n_{\rm th}\sim n_{\rm th}^\star$. Further, we expand all equations to lowest order in $\kappa/\gamma$, while retaining arbitrary orders in $\Phi/\gamma$. After some manipulations, these approximations allow us to derive recurrence relations for the populations $\mu_S^{n,0}$ and $\mu_T^{n,0}$ only, which can be solved in terms of regular continued fractions. We obtain  
\begin{eqnarray}\label{Eq:Effective_Singlet}
		\rho^{\rm ss}_{S}&=&1-\Upsilon\left(\frac{\Phi_{\rm eff}}{\gamma}\right)\left(1+\frac{16\kappa \Phi}{\gamma^2}\right),\\
		\rho^{\rm ss}_{T}&=&\frac{8\kappa \Phi}{\gamma^2}\Upsilon\left(\frac{\Phi_{\rm eff}}{\gamma}\right), \label{Eq:Effective_Triplet}
	\end{eqnarray}
where we introduced the renormalized photon flux $\Phi_{\rm eff}=\Phi \gamma^2/(\gamma^2+24\kappa \Phi)$. When $\Phi\rightarrow\infty$, this parameter saturates at $\Phi_{\rm eff}=\gamma^2/(24\kappa)$.

Based on these expressions for the singlet and triplet populations, we can again use Eq.~\eqref{Eq:Concurrence} to derive an analytic expression for the concurrence for our system,
\begin{equation}\label{Eq:ConcurrenceEffective}
	\mathcal{C}(\rho_{\rm q}^{\rm ss}) = 1-\Upsilon\left(\frac{\Phi_{\rm eff}}{\gamma}\right)\left(1+\frac{24 \kappa \Phi}{\gamma^2}\right).
\end{equation}
In the limit $\kappa\rightarrow 0$, we recover the results from the quasistatic approximation. However, in contrast to all our previous estimates for the concurrence, this expression exhibits a maximum at a finite thermal occupation number, which is approximately given by
\begin{equation}
	\label{Eq:ThermalCritic}
	n_{\rm th}^{\star}\simeq \frac{\gamma^2}{12\kappa^2}.
\end{equation}
At this optimal thermal photon number, the concurrence reaches a value of  
\begin{equation}\label{Eq:MaxConc}
    \mathcal{C}(n_{\rm th}^{\star})=1-2\Upsilon\left(\frac{\gamma}{48\kappa}\right).
\end{equation}
In Fig.~\ref{Fig3:Populations}(a), we compare the predictions for $\mathcal{C}(\rho_{\rm q}^{\rm ss})$ in Eq.~\eqref{Eq:ConcurrenceEffective} and for $n_{\rm th}^{\star}$ in Eq.~\eqref{Eq:ThermalCritic} with the exact numerical results. We find a very good agreement for values of $n_{\rm th}$ up to and beyond $n_{\rm th}^\star$. For even higher thermal occupation numbers, also the state $|11\rangle$ becomes significantly populated [see, for example, the inset in Fig.~\ref{Fig3:Populations}(a)], which is no longer captured by our three-level approximation.

\section{Applications}\label{Sec:Implementations}
To illustrate potential applications for this thermal entanglement mechanism, in this section we discuss two experimental setups where it can be directly employed for the realization of purely passive entanglement distribution schemes. Specifically, we focus on (i) a network of superconducting qubits connected by a microwave quantum channel and (ii) a set of solid-state spin qubits coupled to a phononic waveguide. In both cases the qubit frequencies are typically in the range of a few GHz, which means that when driven by a photon source at room temperature with $T\approx\SI{300}{\kelvin}$, the number of thermal photons $n_{\rm th}\gtrsim 1000$ is already very high and a significant amount of steady-state entanglement is expected. The analysis presented in this section also demonstrates that the entanglement mechanism is robust with respect to various sources of imperfections and that it can be implemented as well in systems with purely bidirectional qubit-waveguide interactions.

\subsection{Superconducting quantum links}\label{Sec:Imp_SC}
As a direct realization of our model in Sec.~\ref{Sec:ThermalWaveguide}, we consider a cryogenic quantum link with two superconducting qubits coupled to a microwave transmission line held at a base temperature of $T_0 \lesssim \SI{50}{\milli\kelvin}$~\cite{Magnard2020,Yam2025}. For this platform, different schemes to realize directional qubit-waveguide couplers have already been implemented, and for the following estimates we consider a setup similar to the one described in Refs.~\cite{Kannan22,Joshi2023,Mirhosseini24,Pfaff25} with a coupling rate of $\gamma/(2\pi)=\SI{10}{\mega\hertz}$. At its input, the microwave channel is coupled to a resistive element, which will emit Johnson-Nyquist noise at temperature $T$. The source and the qubits are separated by a symmetric filter cavity with a total decay rate of $2\kappa=\omega_{\rm c}/Q$. For frequencies $\omega_{\rm q}\simeq \omega_{\rm c}=2\pi\times 5\,$ GHz and realistic quality factors of $Q=2.5\times 10^5$, we obtain $\kappa/\gamma=10^{-3}$ and $n_{\rm th}=1220$ at $T=\SI{293}{\kelvin}$. To describe a realistic scenario, we also account for pure dephasing of the qubits by including the additional dissipator $\frac{\gamma_\phi}{2}\sum_{i}\mathcal{D}[\sigma_i^z]\rho$ in Eq.~\eqref{Eq:ME}. Here, $\gamma_{\phi}=1/T_{\phi}$ is the dephasing rate.

\begin{figure}
		\centering	
        \includegraphics[width=\columnwidth]{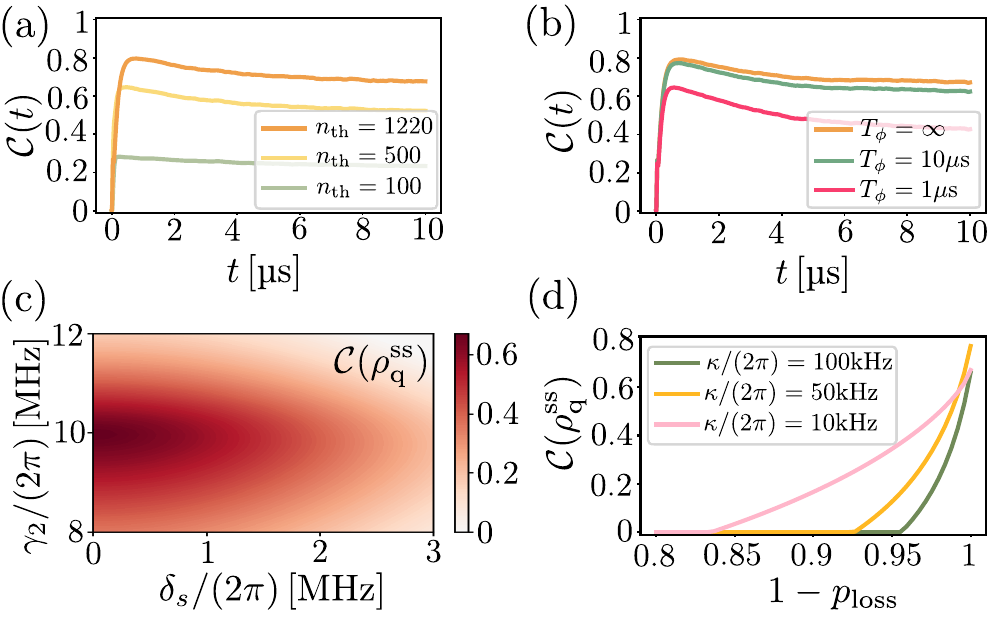}
		\caption{Time evolution of the concurrence $\mathcal{C}(t)\equiv\mathcal{C}(\rho_{\rm q}(t))$ for (a) different thermal photon numbers $n_{\rm th}$ and (b) for a room temperature source with $n_{\rm th}=1220$ for different dephasing times $T_{\phi}$. (c) Steady-state concurrence $\mathcal{C}(\rho^{\rm ss}_{\rm q})$ for  $n_{\rm th}=1220$ and for varying $\gamma_2$ and detunings $\Delta_1=\Delta_2\equiv\delta_s$. (d) Effect of waveguide losses on the steady-state concurrence $\mathcal{C}(\rho^{\rm ss}_{\rm q})$ for different bandwidths $\kappa$ and for $n_{\rm th}=1220$. Unless stated otherwise, we have assumed $\kappa/(2\pi)=\SI{10}{\kilo\hertz}$, $\gamma_i/(2\pi)=\SI{10}{\mega\hertz}$, and $\Delta_i=0$ in all simulations.}
		\label{Fig4:RoomTemperature}
	\end{figure}
    
In Fig.~\ref{Fig4:RoomTemperature}(a), we plot the evolution of the concurrence, assuming that both qubits are initially prepared in state $|0\rangle$. Importantly, this plot shows that while the system takes a long time to relax into the final steady state, the maximum of the concurrence is already reached at a time $\sim\gamma^{-1}$. Therefore, $\gamma^{-1}$, and not $\kappa^{-1}$, is the relevant timescale over which the qubits must stay coherent. This feature is also confirmed by the plot in Fig.~\ref{Fig4:RoomTemperature}(b), where the evolution of $\mathcal{C}$ is shown for different $T_{\phi}$.

Apart from qubit dephasing, there are many other imperfections that can degrade the achievable entanglement in a real experiment. In Fig.~\ref{Fig4:RoomTemperature}(c), we plot the behavior of the steady-state entanglement for scenarios with an asymmetric coupling to the waveguide with $\gamma_2\neq \gamma_1$ and a common detuning of both qubits from the cavity resonance, $\Delta_1=\Delta_2=\delta_s\neq 0$. Note that a purely asymmetric detuning with $\Delta_1=-\Delta_2 = \delta_a \neq 0$ has no significant effect on the entanglement (see discussion in Sec.~\ref{Sec:Imp_Phon} below) and is therefore not considered in this analysis.  Again, we find that the entangling mechanism remains robust, as long as those deviations from the ideal model are small compared to the dominant rate $\gamma$. Finally, in Fig.~\ref{Fig4:RoomTemperature}(d) we investigate the effect of losses in the waveguide that connects the two qubits. Losses can be included in our model by replacing the cascaded interaction between the two qubits with the Liouville operator
\begin{equation}
	\begin{split}
		\mathcal{L}_{\rm casc}^\prime\rho=\gamma\sqrt{1-p_{\rm loss}}\left([\sigma_1^-\rho,
		\sigma_2^+]+[\sigma_2^-,\rho \sigma_1^+]\right),
	\end{split}
\end{equation}
where $p_{\rm loss}$ is the total probability that a photon is lost when propagating from the first to the second qubit. We see that realistic losses of several percent are still acceptable to retain a substantial amount of entanglement and would permit, for example, the use of commercial circulators to realize the unidirectional interactions. Overall, these numerical studies demonstrate that the thermal entanglement mechanism is rather robust and does not rely on a precise fine-tuning of any of the system parameters.

\subsection{Phononic quantum channels}\label{Sec:Imp_Phon}
Another natural setting for the thermal entanglement distribution scheme is a phononic quantum network~\cite{Habraken2012,Gustafsson2014,Schuetz2015,Lemond2018,Bienfait2019,Dumur2021}, where qubits are coupled to propagating acoustic rather than electromagnetic waves. More specifically, in the following we consider a small network of spin qubits associated with silicon-vacancy (SiV) centers in diamond~\cite{Hepp2014}, which are coupled to the quantized acoustic modes of a one-dimensional phononic waveguide via strain interactions. We refer the reader to Refs.~\cite{Lemond2018,Meesala2018,Kuzyk2018,Maity2020,Neuman2021,Arrazola2024} and references therein for more details about potential realizations of such phononic quantum networks.

\begin{figure}
	\centering	\includegraphics[width=\columnwidth]{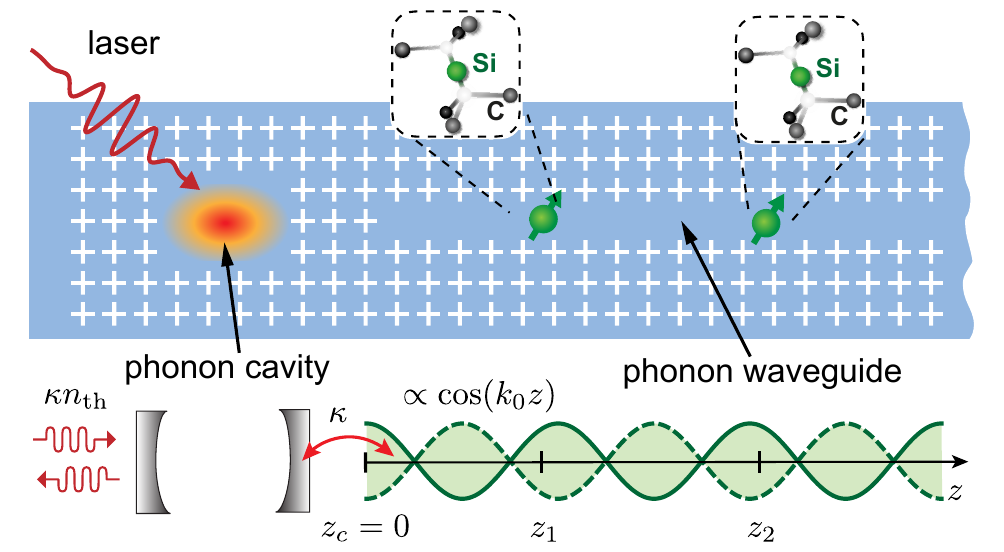}
	\caption{Sketch of a phononic quantum network considered in Sec.~\ref{Sec:Imp_Phon}. Here, phonon cavities and phonon waveguides are realized by defects in an otherwise periodic acoustic bandgap material. The cavity mode can be locally heated by a focused laser beam and emit thermal radiation into the waveguide, which couples to the SiV spin qubits via strain interactions. See text for more details.}
	\label{Fig5:PhononsSetup}
\end{figure}

As depicted in Fig.~\ref{Fig5:PhononsSetup}, the otherwise cold phononic channel can be connected to a hot reservoir via a narrow restriction, which defines an isolated mechanical mode with frequency $\omega_{\rm c}$ and a realistic mechanical quality factor of $Q\gtrsim 10^6$. This mode can be coupled to a second hot phononic channel or it can be artificially heated, for example, through optomechanical interactions~\cite{Vanner2013,shaniv2023thermal}. In this way, the effective local temperature can exceed room temperature or even the melting temperature of the material and much higher thermal occupation numbers of $n_{\rm th}> 10^4$ are possible. However, compared to microwave circuits, directional couplers between the spins and the channel are much more difficult to realize for acoustic waves~\cite{Habraken2012,Lemonde2019}. Therefore, for the following discussion, we will consider a bidirectional waveguide QED system as sketched in Fig.~\ref{Fig5:PhononsSetup}, where the filter cavity is located at the reflecting boundary at $z_{\rm c}=0$ and the qubits located at positions $z_i>0$ are coupled to reflected waves around a resonant wavevector $k_0$. The derivation of the full master equation for this bidirectional, semi-infinite waveguide is outlined in Appendix~\ref{App:Bidirectional}.

\subsubsection{Optimally placed qubits}

By considering, in a first step,  the ideal configuration where all qubits are located at the nodes of the standing wave, i.e. $k_0z_i= 2\pi n_i$ with $n_i=0,1,2,...$, the master equation for this setup reads  
\begin{equation}\label{Eq:FullMEBidirectional}
	\begin{split}
		\dot{\rho}=&-i[H_{\mathrm{1}}+H_2,\rho] +\mathcal{D}[\sqrt{\kappa}a+\sqrt{\gamma}\sigma_1^-+\sqrt{\gamma}\sigma_2^-]\rho\\
		&+\kappa (n_{\mathrm{th}}+1) \mathcal{D}[a]\rho+\kappa n_{\mathrm{th}} \mathcal{D}[a^\dagger]\rho.
	\end{split}
\end{equation}
We see that in this configuration the qubits and the cavity interact only dissipatively and via a fully symmetric decay channel. The cascaded Hamiltonian $H_{\rm casc}$, which was responsible for the existence of the dark state in Eq.~\eqref{Eq:StaticPureState}, is absent. Therefore, in order to restore this coupling between the singlet and the triplet state, we introduce a finite asymmetric detuning $\Delta_1=-\Delta_2=\delta_a$ [see Fig.~\ref{Fig2:Bandwidth}(a)]. In this case, the adiabatic dark state for a static classical field is given by~\cite{Zoller12,Pichler15,Mirhosseini24}
	\begin{equation}\label{Eq:StaticPureState_Phonon}
		|\Psi(\alpha_0)\rangle = \frac{i \delta_a|00\rangle +  \sqrt{2\kappa\gamma}\alpha_0 |S\rangle}{\sqrt{\delta_a^2+2\kappa\gamma |\alpha_0|^2}}.  
	\end{equation}
Apart from a phase, this state has the same structure as the one in Eq.~\eqref{Eq:StaticPureState} and by setting $\delta_a=\gamma/2$ and averaging over $\alpha_0$ we obtain the same adiabatic state as discussed in Sec.~\ref{Sec:Quasistatic}.

\begin{figure}[t]
	\centering	\includegraphics[width=\columnwidth]{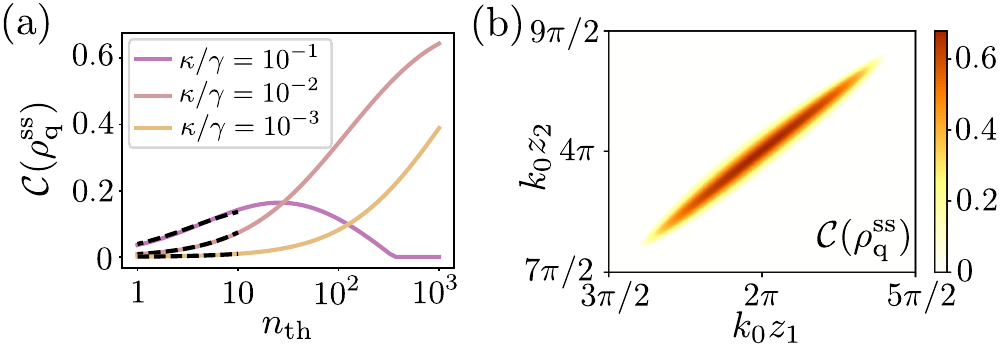}
	\caption[Steady-state entanglement in a phononic network]{Steady-state entanglement between two qubits in a bidirectional phononic quantum network. In (a), the concurrence is plotted as a function of the thermal occupation number for different bandwidth ratios $\kappa/\gamma$ and for qubits located at positions $k_0z_1=2\pi$ and $k_0 z_2=4\pi$.  In (b), the concurrence is plotted for varying positions $z_1$ and $z_2$ for $n_{\mathrm{th}}=2000$ and $\kappa/\gamma=0.01$. The results in both plots are obtained for a fixed $\delta_a=\gamma/2$ and using the continued fraction method described in Appendix~\ref{AppSec:ContinuedFraction}, which neglects the backaction of the qubits on the cavity. In (a), the dashed black lines indicate the results obtained from an exact simulation of Eq.~\eqref{Eq:FullMEBidirectional}.}
	\label{Fig7:PhononsEntanglement}
\end{figure}

Different from a unidirectional channel, the qubits in the phonon waveguide act back on the cavity mode and our theoretical analysis, which was based on an independently evolving thermal field, is in general no longer valid. However, for large $n_{\rm th}$, it is expected that the influence of the qubits on the cavity mode is negligible compared to its coupling to the high-temperature environment. In Fig.~\ref{Fig7:PhononsEntanglement}(a), we verify this intuition by comparing the steady-state entanglement obtained from our phase-space approach with exact numerical simulations of Eq.~\eqref{Eq:FullMEBidirectional} under the same conditions. Over the range of $n_{\rm th}$ over which exact simulations are possible, there is a perfect agreement between the results. This indicates that our numerical methods and analytic estimates discussed in Sec.~\ref{Sec:ContinuedFraction} apply to bidirectional waveguides as well. Note, however, that this is not necessarily the case for larger $\kappa/\gamma\gtrsim 1$. Moreover, for configurations with $k_0 z_i\neq 2\pi n_i$, where more complicated interference effects involving the cavity mode can occur. In this case, while our model somewhat overestimates the amount of entanglement, it accurately captures the dependence on the positions of the qubits.

\subsubsection{General configurations}

Finally, we investigate the case of arbitrary qubit positions $z_i$, which in a bidirectional waveguide affects both the coherent and the incoherent interactions between the qubits and the cavity. In this general case, the equation of motion for the two-qubit operator reads (see Appendix~\ref{App:Bidirectional})
\begin{equation}\label{Eq:StochasticMEPhonons}
	\dot{\mu}(t)=-i[H_\alpha^\prime(t),\mu(t)]+\gamma\mathcal{D}[L_{\mathrm{cos}}]\mu(t),
\end{equation}
where $L_{\mathrm{cos}}=\sum_{i} \cos(k_0z_i) \sigma_i^-$ is the collective jump operator when including the reflection of the emitted waves at $z=0$ and 
\begin{equation}
	\begin{split}
	H_\alpha^\prime(t)=&\sum_i \frac{\Delta_i^\prime}{2}\sigma_i^z  +H^{\prime}_{\mathrm{dip}}\\
    &+i\frac{\sqrt{\kappa \gamma}}{2}[\alpha^*(t) \Sigma-\alpha(t) \Sigma^\dagger].
	\end{split}
\end{equation} 
Here we have defined a position-dependent, collective spin operator $\Sigma=e^{- i k_0 z_1}\sigma^-_1+e^{- i k_0 z_2}\sigma^-_2$, a position-dependent dipole-dipole interaction between the qubits,
\begin{equation}
	H^{\prime}_{\mathrm{dip}}=\frac{J_{12}}{4}(\sigma_1^+\sigma_2^-+\sigma_2^+\sigma_1^-),
\end{equation}
where $J_{12}=\gamma\left(\sin{(k_0(z_1+z_2))}+\sin{(k_0|z_1-z_2|)}\right)$, and a position-dependent shift of the qubit detunings, $\Delta^\prime_i=\Delta_i + J_{ii}$, where $J_{ii}=\gamma_i\sin{(2 k_0 z_i)}/4$.

We see that for arbitrarily placed qubits in a bidirectional channel, interference effects induce several modifications, which are incompatible with the ideal quasistatic dark state. This includes frequency offsets $\Delta^\prime_i\sim \sin(k_0z_i)$ and an additional flip-flop interaction $H_{\rm dip}\sim (\sigma_1^+\sigma_2^-+\sigma_2^+\sigma_1^-)$. Note that in contrast to $H_{\rm casc}$, the dipole-dipole interaction $H_{\rm dip}$ is diagonal in the triple-singlet basis and describes an energy offset rather than a coupling between $|S\rangle$ and $|T\rangle$. To investigate the dependence of steady-state entanglement on those corrections, we evaluate in Fig.~\ref{Fig7:PhononsEntanglement}(b) the concurrence $\mathcal{C}(\rho^{\rm ss}_{\rm q})$ as a function of the positions $z_1$ and $z_2$. We observe that a high value of entanglement is generated in the constrained regions where both $k_0z_1$ and $k_0z_2$ are integer multiples of $\pi$. Note that the same conclusion would hold for a bidirectional superconducting quantum network, in which case the positioning of the qubits within a small fraction of the wavelength can be readily achieved.

\subsubsection{Parameters}
For a specific estimate, we first consider the basic setup discussed in Ref.~\cite{Lemond2018}, where two orbital states of the SiV center with a splitting of $\omega_{\mathrm q, i}/(2\pi)\approx\SI{50}{\giga\hertz}$ are coupled to a narrow phonon waveguide in diamond with a decay rate of $\gamma/(2\pi)\approx 1$ MHz and a bare dephasing time of about $T_\phi\approx 10\,\mu$s.  For this example, a temperature of $T=300$ K translates into $n_{\rm th}\approx 100$ and we obtain a steady-state concurrence of about $\mathcal{C}(\rho_{\rm q}^{\rm ss})\approx 0.2$. Alternatively, as discussed in Refs.~\cite{Meesala2018,Arrazola2024}, one can couple spin states with a much lower splitting of $\omega_{\mathrm q, i}/(2\pi)\approx\SI{3}{\giga\hertz}$ to the propagating acoustic modes of a phononic crystal waveguide~\cite{Joe2024}. 
For this configuration, we assume a lower effective decay rate of $\gamma/(2\pi)\approx 250$ kHz, but at the same time much longer coherence times of $T_\phi\approx 1-10$ ms~\cite{Sukachev2017,Arrazola2024} can be achieved through dynamical decoupling. Under these conditions and for a locally heated source with $n_{\mathrm{th}}\approx  10^3$, our analysis predicts a steady-state entanglement around $\mathcal{C}(\rho_{\rm q}^{\rm ss})\approx 0.62$. These values are mainly limited by the ratio of $\kappa/\gamma$ and can potentially be further increased by engineering phononic structures with higher spin-phonon coupling or higher mechanical quality factors for the filter cavity.

\section{Conclusions}\label{Sec:Conclusion}
In summary, we have presented a fully autonomous scheme for generating long-distance entanglement using thermal photons or phonons as the sole resource. We have shown that the underlying mechanism relies on the existence of an entangled dark state, which is largely insensitive to the precise phase and amplitude of the driving field. This state is adiabatically followed by the system once the bandwidth of the source is sufficiently low, enabling the use of a fully chaotic field to prepare the qubits in an almost pure and highly entangled state. This behavior is in stark contrast to what is observed under the same conditions with a Markovian thermal source. Further, we have developed a general theoretical framework to study the system under conditions of strong thermal driving and have shown that non-adiabatic amplitude fluctuations eventually lead to the breakdown of the entangled state in the limit of very high thermal occupation numbers.

Our analysis of both superconducting and phononic quantum networks shows that this entanglement scheme can also have practically relevant applications. In particular, it enables a purely passive preparation of entangled pairs of superconducting qubits using, for example, the filtered Johnson-Nyquist noise of a resistor at room temperature. Similar to entanglement preparation schemes with coherent fields~\cite{Zoller12,Pichler15}, the mechanism described here for two qubits can be readily generalized for the generation of genuine multipartite entangled states of multiple qubits located along the waveguide, providing an equally simple route for scalability. 

Beyond quantum communication applications, the mechanism described in this work is also of interest in the context of quantum thermodynamics, since it describes the typical setup of a quantum system coupled to a high- and a low-temperature thermal reservoir. The purity or entropy of the system's steady state is, however, mainly determined by the bandwidth of the reservoirs, which provides an additional tuning knob to optimize quantum thermal machines. It will therefore be interesting to explore such thermodynamic applications for this mechanism in the future. 

Some of the numerical simulations used the QuTiP library~\cite{Qutip}. The code used to produce all results in this paper is openly available in Ref.~\cite{Zenodo}.
	
\section*{Acknowledgments}
We thank Yuri Minoguchi and Janine Splettst\"osser for stimulating discussions. This research is part of the Munich Quantum Valley, which is supported by the Bavarian state government with funds from the Hightech Agenda Bayern Plus. J. A. also acknowledges support from the QUANTERA project MOLAR with reference PCI2024-153449, funded by MICIU/AEI/10.13039/501100011033 and the European Union.

\appendix

\section{Bourret approximation}
	\label{Sec:Bourret}
Starting from the master equation for $\mu(\alpha(t),t)\equiv\mu(t)$ in Eq.~\eqref{Eq:StochasticME} and setting $\mathcal{L}_{\alpha}(t)=\alpha(t)\mathcal{L}_++\alpha^*(t)\mathcal{L}_-$, a formal solution of this equation is 
	\begin{equation}\label{eq:DysonSeries}
		\begin{split}
			&\mu(t)=e^{\mathcal{L}_{\rm q} t}\mu(0) + \int_0^t \mathrm{d}t_1 \, e^{\mathcal{L}_{\rm q} (t-t_1)}  \mathcal{L}_\alpha(t_1) e^{\mathcal{L}_{\rm q} t_1} \mu(0)\\
			&+\int_0^t \mathrm{d}t_1\int_0^{t_1} \mathrm{d}t_2\,e^{\mathcal{L}_{\rm q} (t-t_1)}  \mathcal{L}_\alpha(t_1) e^{\mathcal{L}_{\rm q} (t_1-t_2)} \mathcal{L}_\alpha(t_2) \mu(t_2).
		\end{split}
	\end{equation}
    We can average this result over the stochastic trajectories  $\alpha(t)$ to obtain $\rho_{\rm q}(t)=\langle \mu(t)\rangle_{\rm st}$. Since $\langle \alpha(t)\mu(0)\rangle_{\rm st}= \langle \alpha(t)\rangle_{\rm st}\langle \mu(0)\rangle_{\rm st} =0$, the first integral in Eq.~\eqref{eq:DysonSeries} vanishes and after taking the time derivative we obtain
	\begin{equation}
		\begin{split}
			&\dot \rho_{\rm q}(t)=\mathcal{L}_{\rm q} \rho_{\rm q}(t)+ \int_0^{t} \mathrm{d}t'\, \langle \mathcal{L}_\alpha(t) e^{\mathcal{L}_{\rm q} (t-t')} \mathcal{L}_\alpha(t') \mu(t')\rangle_{\rm st}.
		\end{split}
	\end{equation}
In a final step, we make a Bourret or decorrelation approximation by factorizing expectation values as  
	\begin{equation}
		\langle \alpha(t)\alpha^*(t')\mu(t')\rangle_{\rm st} \simeq  \langle \alpha(t)\alpha^*(t')\rangle_{\rm st} \rho_{\rm q}(t').
	\end{equation}
After this approximation and since $\langle \alpha(t)\alpha(t') \rangle_{\rm st}=0$, we end up with Eq.~\eqref{Eq:WeakME} in the main text. 
	
We can use the Laplace transform to solve this time-nonlocal equation. By defining $\widehat \rho_{\rm q}(s)=\int_0^\infty \mathrm{d}t' \, e^{-st'} \rho_{\rm q}(t')$, we obtain 
\begin{equation}
	\begin{split}
		s \widehat\rho_{\rm q}(s) = \,&\rho_{\rm q}(0)+ \mathcal{L}_{\rm q} \widehat\rho_{\rm q}(s)\\
		&+\frac{n_{\rm th}}{2}\mathcal{L}_+ [(s+\kappa)\mathcal{I}-\mathcal{L}_{\rm q}]^{-1}\mathcal{L}_- \widehat\rho_{\rm q}(s) \\ 
		&+\frac{n_{\rm th}}{2}\mathcal{L}_- [(s+\kappa)\mathcal{I}-\mathcal{L}_{\rm q}]^{-1}\mathcal{L}_+ \widehat\rho_{\rm q}(s),
	\end{split}
\end{equation}
where $\mathcal{I}$ is the identity operator and we have used that $\langle \alpha(t)\alpha^*(t')\rangle_{\rm st}=n_{\rm th} e^{-\kappa |t-t'|}/2$. By solving this equation for $ \widehat\rho_{\rm q}(s) $ and taking the limit $\rho_{\rm q}(t\rightarrow \infty)={\lim}_{s\rightarrow 0} \,s \widehat\rho_{\rm q}(s) $ we obtain the steady-state density matrix. The only nonvanishing matrix elements are the populations
    \begin{align}
	   \mathcal{N} \rho^{\rm ss}_{1} =\, & 64 \kappa \Phi^2 (\gamma+\kappa), \nonumber\\
	   \mathcal{N} \rho^{\rm ss}_T =\, & 8 \kappa \Phi [(3\gamma+2\kappa)^2+8\Phi(\gamma+\kappa)],\nonumber\\
	   \mathcal{N} \rho^{\rm ss}_S =\, & 8 \Phi[(\gamma+\kappa)(3\gamma+2\kappa)^2+8\Phi(\gamma^2+\kappa^2+\gamma\kappa)],\nonumber\\
	   \mathcal{N} \rho^{\rm ss}_{0} =\, & 64\kappa\Phi^2(\gamma+\kappa)+8\Phi(\gamma+2\kappa)^3 \nonumber\\
       &+(\gamma+2\kappa)^2(3\gamma+2\kappa)^2, \nonumber
    \end{align}
 where $\mathcal{N}=(\gamma+2\kappa)(\gamma+2\kappa+8\Phi) [(3\gamma+2\kappa)^2+8\Phi(\gamma+2\kappa)])$. By keeping only the lowest relevant orders of $\kappa/\gamma$, we obtain the approximate results given in Sec.~\ref{Sec:BourretApproximation}.

\section{Continued fraction methods}\label{AppSec:ContinuedFraction}
In Eq.~\eqref{Eq:FPrho} in Sec.~\ref{Sec:StochasticME} we introduced a phase-space representation of the full system density operator, which can be sampled in terms of stochastic trajectories $\alpha(t)$ and the corresponding equation of motion for $\mu(\alpha(t),t)$ given in Eq.~\eqref{Eq:StochasticME}. This approach is not limited by the size of the Hilbert space, but achieving sufficiently low statistical errors is still computationally demanding. Alternatively, as outlined in Sec.~\ref{Sec:ContinuedFraction}, we can solve the combined equation of motion given in Eq.~\eqref{Eq:FokkerPlankME}, by expanding $\mu(\alpha,t)$ in terms of the eigenfunctions of the Fokker-Planck operator $\Lambda$. In this appendix, we summarize the main results of this approach. 

Our starting point is the expansion of $\mu(\alpha(t),t)$ in Eq.~\eqref{eq:muExpansion}. The equations of motion for the coefficients of this expansion,
\begin{equation}
		\mu^{n,m}(t)=\int \mathrm{d}^2\alpha\, \phi_{n,m}^*(\alpha) \mu(\alpha,t),
\end{equation}
can be derived from integrating both sides of Eq.~\eqref{Eq:FokkerPlankME} over the complex plane and using the orthogonality relation 
\begin{equation}
\int \mathrm{d}^2\alpha\, \phi_{n,m}^*(\alpha) P_{n',m'}(\alpha) = \delta_{n,n'}\delta_{m,m'},
\end{equation}
together with the recurrence relations for the generalized Laguerre polynomials,
\begin{eqnarray*}
L_n^{m-1}(x) &=& L_n^m(x)-L^m_{n-1}(x), \\
x L_n^{m+1}(x)&=& (n+m+1)L_n^m(x)-(n+1)L^m_{n+1}(x). 
\end{eqnarray*}
As a result, we obtain one set of equations for $m=0$,
	\begin{equation}
		\label{Eq:Alpha0}
		\begin{split}
			&\left(\frac{\mathrm{d}}{\mathrm{d}t}+2\kappa n\mathcal{I}\right)\mu^{n,0}=\mathcal{L}_{\rm q}\mu^{n,0}\\
			&+\sqrt{\frac{n_{\rm th}}{2}}\mathcal{L}_+\left(\sqrt{n+1}\mu^{n,1}-\sqrt{n}\mu^{n-1,1}\right)\\
			&+\sqrt{\frac{n_{\rm th}}{2}}\mathcal{L}_-\left(\sqrt{n+1}\mu^{n,-1}-\sqrt{n}\mu^{n-1,-1}\right).
		\end{split}
	\end{equation}
and another set of equations for $m\neq 0$,
	\begin{equation}
		\label{Eq:AlphaP}
		\begin{split}
			&\left[\frac{\mathrm{d}}{\mathrm{d}t}+\kappa (2n+|m|)\mathcal{I}\right]\mu^{n,m}=\mathcal{L}_{\rm q}\mu^{n,m}\\
			&+\sqrt{\frac{n_{\rm th}}{2}}\mathcal{L}_{\pm}\left(\sqrt{n+|m|+1}\mu^{n,m\pm1}-\sqrt{n}\mu^{n-1,m\pm1}\right)\\
			&+\sqrt{\frac{n_{\rm th}}{2}}\mathcal{L}_\mp\left(\sqrt{n+|m|}\mu^{n,m\mp1}-\sqrt{n+1}\mu^{n+1,m\mp1}\right),
		\end{split}
	\end{equation}
where the upper (lower) sign applies for $m>0$ ($m<0$).

\subsection{Numerics}
In this general formulation, we obtain an infinite set of coupled equations for $n$ and $m$. However, for our specific example, we can restrict this hierarchy to values of $m=0,\pm1,\pm 2$. This restriction arises from the fact that our two-qubit system can support at most two excitations and $(\mathcal{L}_\pm)^k=0$ for $k>2$. Further, in the steady state, the operators $\mu^{n,\pm2}$ can be expressed in terms of the other coefficients as
\begin{equation}
\begin{split}
    \mu^{n,\pm2}=&\sqrt{\frac{n_{\rm th}}{2}}\left[(\kappa(2n+2)\mathcal{I}-\mathcal{L}_{\rm q})\right]^{-1} \\
    &\times \mathcal{L}_{\mp}\Big(\sqrt{n+2}\mu^{n,\pm 1}-\sqrt{n+1}\mu^{n+1,\pm 1}\Big).
\end{split}
\end{equation}
The remaining operators can be grouped into vectors $\sigma^n=\left(\mu^{n,0},\mu^{n,1},\mu^{n,-1}\right)^{\rm T}$, which obey the recurrence  relation
\begin{equation}\label{eq:SigmaRecurrence}
		A_n \sigma^n + B_n \sigma^{n-1} + C_n \sigma^{n+1} = 0.
\end{equation}
Here the matrices $A_n$, $B_n$ and $C_n$ are given by
\newpage
\begin{equation}
		A_n=-
		\begin{pmatrix}
			\mathcal{L}_{\rm q}- 2 n \kappa\mathcal{I}  &  \sqrt{\frac{n_{\rm th}(n+1)}{2}}\mathcal{L}_{+}&  \sqrt{\frac{n_{\rm th}(n+1)}{2}}\mathcal{L}_{-}\\
			\sqrt{\frac{n_{\rm th}(n+1)}{2}}\mathcal{L}_{-} & -a_{+} & 0\\
			\sqrt{\frac{n_{\rm th}(n+1)}{2}}\mathcal{L}_{+} & 0 & -a_{-}
		\end{pmatrix},\\
	\end{equation}
	\begin{equation}
		B_n=\sqrt{n}
		\begin{pmatrix}
			0 & \sqrt{\frac{n_{\rm th}}{2}}\mathcal{L}_{+} & \sqrt{\frac{n_{\rm th}}{2}}\mathcal{L}_{-}\\
			0 & \sqrt{n+1}\mathcal{P}^{0}_{+} & 0\\
			0 & 0 & \sqrt{n+1}\mathcal{P}^{0}_{-}
		\end{pmatrix},\\
	\end{equation}
    and
	\begin{equation}
		C_n=\sqrt{n+1}
		\begin{pmatrix}
			0 & 0 & 0\\
			\sqrt{\frac{n_{\rm th}}{2}}\mathcal{L}_{-} & \sqrt{n+2}\mathcal{P}^{2}_{+} & 0\\
			\sqrt{\frac{n_{\rm th}}{2}}\mathcal{L}_{+} & 0 & \sqrt{n+2}\mathcal{P}^{2}_{-}
		\end{pmatrix},
        \end{equation}
	where we have introduced the abbreviations 
	\begin{equation}
		a_{\pm}=\kappa(2n+1)\mathcal{I}-\mathcal{L}_{\rm q}-(n+2)\mathcal{P}^{2}_{\pm}-n\mathcal{P}^{0}_{\pm},
	\end{equation}
	and
	\begin{equation}
		\mathcal{P}^{x}_{\pm}=\frac{n_{\rm th}}{2}\mathcal{L}_{\pm}[\kappa(2n+x)\mathcal{I}-\mathcal{L}_{\rm q}]^{-1}\mathcal{L}_{\mp}.
	\end{equation}
By representing all superoperators as matrices, the recurrence  relation in Eq.~\eqref{eq:SigmaRecurrence} can be solved in terms of a matrix continued fraction~\cite{Risken1996}, 
	\begin{equation}
		\label{Eq:MatrixCF}
		\Big[A_0+\mathcal{K}\Big]\sigma^0=0,
	\end{equation}
    where 
	\begin{equation}
		\mathcal{K}=C_0\dfrac{\mathcal{I}}{-A_1-C_1\dfrac{\mathcal{I}}{-A_2-C_2\dfrac{\mathcal{I}}{-A_3-...}B_2}B_2}B_1.
	\end{equation}
The physical state of interest is then given by the first element of $\sigma^0$. 

Note that to a lowest-order approximation, $\mathcal{K}=0$ and the steady state defined by $A_0\sigma^0=0$ is equivalent to the result obtained from the Bourret approximation~\cite{cirac1991thermal}. More generally, $\sigma^0$ can be solved numerically by truncating the matrix continued fraction at a large enough $n_{\rm max}$. Typically, we find a convergence of the results for $n_{\rm max}\approx \sqrt{n_{\rm th}}$ [see Fig.~\ref{Fig7_Convergence}(a)], which makes this approach numerically very efficient.

\subsection{Three-level approximation}
\begin{figure}[t]
	\centering	\includegraphics[width=\columnwidth]{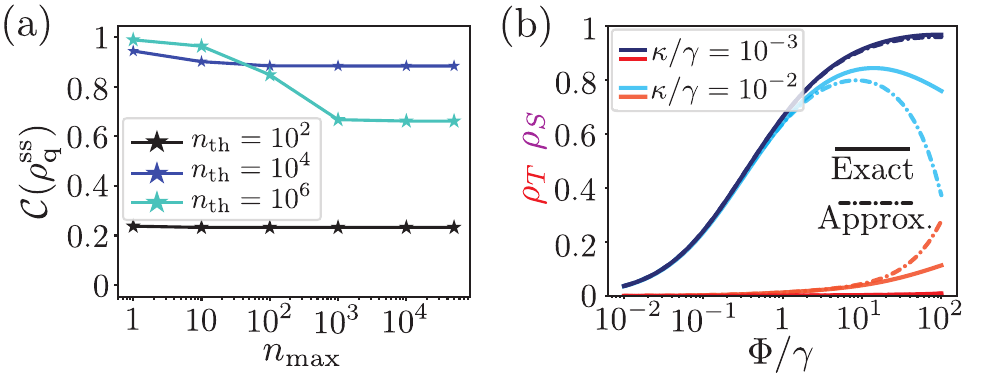}
	\caption{(a) Convergence of the continued fraction introduced in Eq.~\eqref{Eq:MatrixCF}, quantified in terms of the steady-state concurrence and for a typical bandwidth of $\kappa/\gamma=10^{-3}$. (b) Comparison of the singlet and triplet populations obtained from the exact continued fraction in Eq.~\eqref{eq:RecurrenceThreeLevel} (solid lines) with the approximate results obtained from the matrices in Eqs.~\eqref{eq:AppAn_approximation}-\eqref{eq:AppAn_approximation_C} with $\mathcal{E}_n=0$ (dashed-dotted line).}
	\label{Fig7_Convergence}
\end{figure}
To go beyond numerics and derive approximate analytic results, we make a three-level approximation by omitting transitions to the state $|11\rangle$. A numerical justification of this approximation is already given in the inset of Fig.~\ref{Fig3:Populations}(a), where the population $\rho_{11}$ is negligible up to and slightly beyond $n^\star_{\rm th}$. This approximation restricts the possible values of $m$ to $m=0,\pm1$. Further, in the steady state, we can express the coefficients $\mu^{n,\pm1}$ solely in terms of $\mu^{n}\equiv \mu^{n,0}$ and derive a single recurrence relation 
	\begin{equation}\label{eq:RecurrenceThreeLevel}
		a_n \mu^{n}+b_n \mu^{n-1} + c_n \mu^{n+1}=0 
	\end{equation}
with
\begin{flalign}
	&a_n=2\kappa n \mathcal{I}-\mathcal{L}_{\rm q} - \frac{n_{\rm th}}{2}\left[(n+1)\Gamma_{+}(n)+n\Gamma_{-}(n)\right]
\end{flalign}
and
\begin{flalign}
	&b_n=\frac{n_{\rm th}}{2} n \Gamma_-(n),\\
	&c_n=\frac{n_{\rm th}}{2}(n+1)\Gamma_+(n).
\end{flalign}
We have also defined 
\begin{equation}
	\begin{split}
		\Gamma_{\pm}(n)=\,&\mathcal{L}_{+}[\kappa(2n\pm1)\mathcal{I}-\mathcal{L}_{\rm q}]^{-1}\mathcal{L}_{-}\\
		&+\mathcal{L}_{-}[\kappa(2n\pm1)\mathcal{I}-\mathcal{L}_{\rm q}]^{-1}\mathcal{L}_{+}.
	\end{split}
\end{equation}

The three-term recurrence relation in Eq.~\eqref{eq:RecurrenceThreeLevel} can again be solved for $\mu^0$ in terms of a matrix continued fraction. However, to proceed with analytic calculations, we express this equation explicitly in terms of the matrix elements $\mu^n_{i,j}$ with $i,j\in\{0,S,T\}$. We obtain 
\begin{widetext}
\begin{subequations}\label{Eq:EOM}
    \begin{align}
        \left(2\gamma+2\kappa n\right)\mu^n_{T}&=\frac{\gamma}{2}\chi^n_{S,T}-\sqrt{2\gamma \Phi}\left(\sqrt{n+1}\chi^n_{0,T}-\sqrt{n}\chi^{n-1}_{0,T}\right),\\
         2\kappa n \mu^n_{S}&=-\frac{\gamma}{2}\chi^n_{S,T},\\
	    (\gamma+2\kappa n)\chi^n_{S,T}&=\gamma(\mu_{S}^n-\mu_{T}^n)-\sqrt{2\gamma \Phi}\left(\sqrt{n+1}\chi_{0,S}^{n}-\sqrt{n}\chi_{0,S}^{n-1}\right),\\
	    \kappa(2n+1)\chi^n_{0,S}&=-\frac{\gamma}{2}\chi_{0,T}^n+\sqrt{2\gamma \Phi (n+1)}\left(\chi_{S,T}^n-\chi_{S,T}^{n+1}\right),\\
        \left(\gamma+\kappa(2n+1)\right)\chi^n_{0,T}&=\frac{\gamma}{2}\chi_{0,S}^n-2\sqrt{2\gamma \Phi(n+1)}\left(\mu_{0,0}^n-\mu_{0,0}^{n+1}-\mu_{T}^n+\mu_{T}^{n+1}\right),
    \end{align}
\end{subequations}
\end{widetext}
where $\mu^n_{S,T}=\langle S/T|\mu^n|S/T\rangle$ for short and $\chi^n_{i,j}=\mu^n_{i,j}+\mu^n_{j,i}$. The equation for the ground-state population is determined by the trace of $\rho_{\rm q}$, i.e. $\mu_{0,0}^n=\delta_{n,0}-\mu_{S}^n-\mu_{T}^n$.  After eliminating the equations for the coherences, we are left with two coupled recurrence relations for the populations $\mu_S^n$ and $\mu_T^n$. These can be written in a compact form as
\begin{equation}\label{AppEq:CF_SingletTriplet}
    A_n\sigma^n+B_n\sigma^{n-1}+C_n\sigma^{n+1}=Y_0(\delta_{n,0} - \delta_{n-1,0}),
\end{equation}
where we have introduced the two-component vectors $\sigma^n=(\mu_T^n,\mu_S^n)^{\rm T}$ and $Y_0=(-8 \Phi \gamma^2, 16\Phi\gamma\kappa)^{\rm T}/(\gamma+2\kappa)^2$. 
The full expressions of the matrices $A_n$, $B_n$, and $C_n$ are rather lengthy and not presented here. Instead, we focus on the quasistatic limit and expand Eq.~\eqref{AppEq:CF_SingletTriplet} to the lowest relevant order in the parameter $\kappa/\gamma \ll 1$, while keeping all orders of $\Phi/\gamma$. As a result, we obtain
\begin{equation}\label{eq:AppAn_approximation}
    A_n\approx 
        \begin{pmatrix}
            \gamma-16\Phi (2n+1) & -\gamma-8\Phi(2n+1) \\
            2\gamma+ 32\Phi\kappa/\gamma + 4\mathcal{E}_n &   16 \kappa \Phi/\gamma+ \mathcal{E}_n
    \end{pmatrix},
\end{equation}
\begin{equation}
    B_n\approx 
        \begin{pmatrix}
            16\Phi n & 8\Phi n \\
            0 & 0
    \end{pmatrix},
\end{equation}
    \begin{equation}\label{eq:AppAn_approximation_C}
    C_n\approx
        \begin{pmatrix}
            16\Phi (n+1) & 8\Phi (n+1) \\
            0 & 0
    \end{pmatrix},
\end{equation}
and $Y_0\approx(-8 \Phi, 16\Phi \kappa/\gamma)^{\rm T}$. In Eq.~\eqref{eq:AppAn_approximation} we have defined $\mathcal{E}_n=16\Phi\kappa(n+2n^2)/\gamma$, which, in a last approximation, we set to zero. While the validity of this approximation is not immediately obvious, it can be justified, either by an analytic evaluation of the first few orders of the continued fraction or by the comparison with the exact result obtained from Eq.~\eqref{AppEq:CF_SingletTriplet} [see Fig.~\ref{Fig7_Convergence}(b)]. 

After these approximations, Eq.~\eqref{AppEq:CF_SingletTriplet} can be decoupled into two separate recurrence relations for the singlet and the triplet population, which are given by
\begin{equation}
	\label{Eq:SingletEffRecurrence}
	\begin{split}	&\Big(\gamma^2+8\Phi[(2n+1)\gamma+3\kappa]\Big)\rho_{S}^{n}=\\
		&+8n\Phi\gamma\rho_{S}^{n-1}+8(n+1)\Phi\gamma\rho_{S}^{n+1}\\
		&+8\Phi(\gamma+\kappa)\delta_{n,0}-8\Phi(\gamma+3\kappa+48\Phi\kappa^2/\gamma^2)\delta_{n-1,0},
	\end{split}
\end{equation}
and 
\begin{equation}
	\label{Eq:TripletEffRecurrence}
	\begin{split}
&\Big(\gamma^2+8\Phi[(2n+1)\gamma+3\kappa]\Big)\rho_{T}^n =\\
		&+ 8\Phi n \gamma\rho_{T}^{n-1}+ 8(n+1)\Phi \gamma\rho_{T}^{n+1}\\
		&+ 8\Phi\kappa \delta_{n,0}+\frac{3(8\Phi \kappa)^2}{\gamma^2}\delta_{n-1,0},
	\end{split}
\end{equation}
respectively. Both equations have the same structure,  
\begin{equation}
	 a_nX_n= b_n X_{n-1} + c_n X_{n+1}+Y_0\delta_{n,0}+Y_1\delta_{n-1,0}, 
\end{equation}
and a solution given by
\begin{equation}
	\label{Eq:CFGeneral}
	\begin{split}
		X_0&=\frac{\Big(Y_0+\frac{c_0}{a_1-\frac{c_1b_2}{a_2-...}}Y_1\Big)}{a_0-\frac{c_0 b_1}{a_1-c_1\frac{c_1 b_2}{a_2-...}}}=\mathcal{F}_1(Y_0+\mathcal{F}_2Y_1).
	\end{split}
\end{equation}
Here we have introduced the two regular continued fractions
\begin{equation}
	\label{Eq:CF1}
	\mathcal{F}_1=\frac{1}{a_0-\dfrac{c_0 b_1}{a_1-c_1\dfrac{c_1 b_2}{a_2-...}}}
\end{equation}
and
\begin{equation}
	\label{Eq:CF2}
	\mathcal{F}_2=\frac{c_0}{a_1-\dfrac{c_1b_2}{a_2-...}}.
\end{equation}
\subsection{Theory of continued fractions}
After having reduced our solution for the steady-state populations to two continued fractions, we can apply known methods to solve them in terms of their continuants. For a general continued fraction, a closed expression can be found by~\cite{Risken1996}
\begin{equation}
	\mathcal{F}=\frac{b_0}{a_0+\frac{b_1}{a_1+...}}=\lim_{n\rightarrow\infty}\frac{P_n}{Q_n},
\end{equation}
where the so-called continuants fulfill the same relations,
\begin{subequations}
\begin{align}
		& P_{n+2}=a_{n+1} P_{n+1}+b_{n+1} P_{n},\\
		& Q_{n+2}=a_{n+1} Q_{n+1}+b_{n+1} Q_{n},
\end{align}
\end{subequations}
but with different initial conditions $P_0=0$, $Q_0=1$, $P_1=1$, and $Q_1=a_0$. By introducing the corresponding generating functions, $P(z)=\sum_{n\ge 0}P_n z^n$ and $Q(z)=\sum_{n\ge 0}Q_n z^n$, we can relate the asymptotic limit of the coefficients of the continuants to~\cite{Flajolet2009}
\begin{equation}\label{Eq:Series}
	\lim_{n\rightarrow\infty}\frac{P_n}{Q_n}=\lim_{z\rightarrow z^{\star}}\frac{P(z)}{Q(z)},
\end{equation}
where $z^{\star}$ is the radius of convergence of $P(z)$ and $Q(z)$, i.e. its singular point if any.
	
For the continued fraction $\mathcal{F}_1$ given in Eq.~\eqref{Eq:CF1} we obtain 
\begin{align}
		& P_{n+2}=[\gamma^{\prime}+x(2n+3)]P_{n+1}-x^2 (n+1)^2 P_{n},
\end{align}
\noindent with $x=8\Phi\gamma$, $\gamma^{\prime}=\gamma^2+24\kappa \Phi$ and $a_0=\gamma^{\prime}+ x$. We define $P_n=n! p_n$ to transform this recurrence relation into the form
\begin{align}
		&(n+2)p_{n+2}=[\gamma^{\prime}+x(2n+3)]p_{n+1}-x^2 (n+1) p_{n},
\end{align}
which translates into the first-order ordinary differential equation
\begin{equation}
	\begin{split}
		p^{\prime}(z)(1-xz)^2 z=&p(z)[(\gamma^{\prime}+x) z- x^2 z^2]\\
		&-z(\gamma^{\prime}+x)p_0+z p_1
	\end{split}
\end{equation}
for the generating function $p(z)=\sum_{n \ge0}p_n z^n$. Its formal solution is
\begin{equation}
	\begin{split}
		&p(z)=\frac{e^{\frac{z\gamma^{\prime}}{1-xz}}}{x(1-xz)}\Big[x p_0\\
		&+e^{\frac{\gamma^{\prime}}{x}}p_1 \left(\Gamma[0,\gamma^{\prime}/x]+\Gamma[0,\frac{\gamma^{\prime}}{x(xz-1)}]\right)\Big],
	\end{split}
\end{equation}
which is singular at $z=1/x$. The same procedure can be applied for the continuants $Q_n$ with the corresponding transformed generating function $q(z)=\sum_{n \ge0}q_n z^n$, but different initial conditions. Altogether we obtain
	\begin{equation}
		\mathcal{F}_1=\lim_{n\rightarrow\infty}\frac{p_n}{q_n}=\lim_{z\rightarrow 1/x}\frac{p(z)}{q(z)}=\frac{e^{\frac{\gamma^{\prime}}{x}}\Gamma[0,\frac{\gamma^{\prime}}{x}]}{x},
	\end{equation}
where we made use of the fact that $p_n/q_n=P_n/Q_n$.

Similarly, for the continued fraction $\mathcal{F}_2$ given in Eq.~\eqref{Eq:CF2} the continuants fulfill 
\begin{subequations}
\begin{align}
		& P_{n+2}=a_{n+2} P_{n+1}-c_{n+1} b_{n+2} P_{n},\\
		& Q_{n+2}=a_{n+2} Q_{n+1}-c_{n+1} b_{n+2} Q_{n},
\end{align}
\end{subequations}
with initial conditions $P_0=0$, $Q_0=1$, $P_1=c_0=x$, and  $Q_{1}=a_1=\gamma^{\prime}+3x$. In this case, we define $P_n=(n+1)! p_n$ to obtain the transformed recurrence relation 
\begin{subequations}
\begin{align}
		& (n+3)p_{n+2}=[\gamma^{\prime}+x(2n+5)]p_{n+1}-x^2 (n+2) p_{n}.
\end{align}
\end{subequations}
The corresponding differential equation for the generating function,
\begin{equation}
	\begin{split}
		p^{\prime}(z)(1-x z)^2z=&p(z)[z(\gamma^{\prime}+3x)-2x^2z^2-1]\\
		&+(1-z(\gamma^{\prime}+3x))p_0+2z p_1,
	\end{split}
\end{equation}
has a similar structure to the above and we can follow the same steps to obtain
\begin{equation}
	\mathcal{F}_2=1+\frac{\gamma^{\prime}}{x}-\frac{e^{-\gamma^{\prime}/x}}{\Gamma[0,\gamma^{\prime}/x]}.
\end{equation}

Based on these closed expressions for $\mathcal{F}_1$ and $\mathcal{F}_2$ and the general solution given in Eq.~\eqref{Eq:CFGeneral}, we obtain the singlet population 
\begin{equation}
	\label{Eq:EffSinglet}
	\rho_{S}=1-\Big[\Upsilon(\Phi_{\rm eff}/\gamma)\left(1+\frac{3\kappa}{\gamma}\right)-\frac{3\kappa}{\gamma}\Big]\left(1+\frac{16\kappa \Phi}{\gamma^2}\right),
\end{equation}
with $\Phi_{\rm eff}=\Phi \gamma^2/(\gamma^2+24\kappa \Phi)$, and 
\begin{equation}
	\label{Eq:EffTriplet}
	\rho_{T}=\frac{8\kappa \Phi}{\gamma^2}[\Upsilon(\Phi_{\rm eff}/\gamma)(1+3\kappa/\gamma)-3\kappa/\gamma],
\end{equation}
for the triplet state. By keeping only the lowest order in $\kappa/\gamma$, we obtain Eq.~\eqref{Eq:Effective_Singlet} and Eq.~\eqref{Eq:Effective_Triplet} in the main text.

\section{Semi-infinite waveguide}\label{App:Bidirectional}
In this appendix, we briefly outline the derivation of the master equation for a semi-infinite waveguide. 
This scenario can be modeled by a bidirectional qubit-waveguide interaction of the form
\begin{equation}\label{eq:AppCHint}
    H_{\rm int}=2i\sum_{j=0}^2 g_j \int \mathrm{d}\omega  \, b^\dagger(\omega)c_j \cos{(\omega z_j/v)} - \mathrm{H.c.},
\end{equation}
where $b(\omega)$ are bosonic annihilation operators that fulfill $[b(\omega),b^\dagger(\omega^{\prime})]=\delta(\omega-\omega^{\prime})$. In Eq.~\eqref{eq:AppCHint}, $c_0=a$ and $c_{i>0}=\sigma^-_{i>0}$ denote the jump operators for the filter cavity and the two qubits, respectively. In the case of the phonon waveguide considered in Sec.~\ref{Sec:Imp_Phon}, the waveguide mode represents the displacement field with a maximum at the position $z_0\equiv z_c=0$ of the cavity. However, waveguide setups with different boundary conditions and coupling mechanisms can also be treated within the same framework by simply adjusting the positions $z_i$ accordingly. For example, for qubits coupled to strain, i.e., the gradient of the displacement field, the $z_{i=1,2}$ in Eq.~\eqref{eq:AppCHint} will differ from the actual positions by $\Delta z_i= \pi/(2k_0)$. 

By assuming linear dispersion relations and the validity of the usual Markovian approximation, we can use the standard approach~\cite{Zoller04} to eliminate the waveguide modes and derive a master equation for the cavity and the qubits only. This master equation is of the general form (see, for example, Refs.~\cite{Pichler2016,Fayard2021}) 
\begin{equation}\label{AppEq:PhononME}
\begin{split}
    \dot{\rho}=&-i[H_1+H_2+\sum_{j,l}J_{j,l}c_j^\dagger c_l,\rho]+\sum_{j,l}\Gamma_{j,l}\mathcal{D}[c_j,c_l]\rho\\
    &+\kappa(n_{\rm th}+1)\mathcal{D}[a]\rho+\kappa n_{\rm th}\mathcal{D}[a^\dagger]\rho.
\end{split}
\end{equation}
Here, we have the coherent coupling $J_{j,l}=\sqrt{\gamma_j\gamma_l}\left(\sin{(k_0(z_j+z_l))}+\sin{(k_0|z_j-z_l|)}\right)/4$, as well as the decay rates $\Gamma_{j,l}= \sqrt{\gamma_j\gamma_l}\left(\cos{(k_0(z_j+z_l))}+\cos{(k_0|z_j-z_l|)}\right)/2$ and we have introduced the short notation $\mathcal{D}[c_j,c_l]\rho=c_j\rho c_l^\dagger-\{c_l^\dagger c_j,\rho\}/2$. Note that by identifying $\gamma_0\equiv \kappa$, Eq.~\eqref{AppEq:PhononME} is defined such that the cavity and each qubit individually decay with the rates $\kappa$ and $\gamma_i\cos^2(k_0z_i)$, respectively. This allows us to make a direct comparison with the unidirectional setup. 

When introducing the phase-space representation as in Sec.~\ref{Sec:StochasticME}, the master equation can be mapped onto a partial differential equation using the substitution rules for the P-function~\cite{Zoller04},
\begin{subequations}
\begin{align}
    &a \rho \rightarrow \alpha P(\alpha),\\
    &a^\dagger \rho \rightarrow \left(\alpha^*-\partial_\alpha\right) P(\alpha),\\
    & \rho a^\dagger \rightarrow \alpha^*  P(\alpha),\\
    & \rho a \rightarrow \left(\alpha-\partial_{\alpha^*}\right) P(\alpha).
\end{align}
\end{subequations}
In contrast to the cascaded setup discussed in Sec.~\ref{Sec:StochasticME}, in a bidirectional waveguide, this procedure will also introduce additional terms of the form
\begin{equation}
    -i\sum_jJ_{0,j}\left\{[\alpha^*\sigma_j^-+\alpha\sigma^+_j,\mu]-\partial_\alpha \sigma^-_j\mu + \partial_{\alpha^*}\mu \sigma_j^+\right\},
\end{equation}
and
\begin{equation}
    \sum_j \frac{\Gamma_{0,j}}{2}\left\{[\alpha^*\sigma_j^--\alpha\sigma^+_j,\mu]+\partial_{\alpha^*}\mu \sigma_j^++\partial_{\alpha} \sigma_j^-\mu\right\}.
\end{equation}
For the numerical simulations presented in Sec.~\ref{Sec:Imp_Phon}, the terms proportional to $\partial_\alpha$ and $\partial_{\alpha^*}$, which describe the backaction of the qubits on the cavity, are neglected.

\bibliographystyle{quantum}

\end{document}